\documentclass[]{ws-rv9x6} 
\usepackage{ws-rv-har}    
\usepackage{ws-rv-thm}    
\usepackage{subfigure}    
\usepackage{enumitem}
\usepackage[percent]{overpic}
\usepackage{xcolor}
\makeindex

\begin{document}		
\newcommand{\ltsima}{$\; \buildrel < \over \sim \;$}
\newcommand{\lsim}{\lower.5ex\hbox{\ltsima}}
\newcommand{\gtsima}{$\; \buildrel > \over \sim \;$}
\newcommand{\gsim}{\lower.5ex\hbox{\gtsima}}
\newcommand{\bra}{\langle}
\newcommand{\ket}{\rangle}
\newcommand{\lprime}{\ell^\prime}
\newcommand{\lpp}{\ell^{\prime\prime}}
\newcommand{\mprime}{m^\prime}
\newcommand{\mpp}{m^{\prime\prime}}
\newcommand{\ci}{\mathrm{i}}
\newcommand{\dd}{\mathrm{d}}
\newcommand{\veck}{\mathbf{k}}
\newcommand{\vecx}{\mathbf{x}}
\newcommand{\vecr}{\mathbf{r}}
\newcommand{\vecv}{\mathbf{\upsilon}}
\newcommand{\vecw}{\mathbf{\omega}}
\newcommand{\vecj}{\mathbf{j}}
\newcommand{\vecq}{\mathbf{q}}
\newcommand{\vecl}{\mathbf{l}}
\newcommand{\vecn}{\mathbf{n}}
\newcommand{\lm}{\ell m}
\newcommand{\that}{\hat{\theta}}
\newcommand{\thatp}{\that^\prime}
\newcommand{\chip}{\chi^\prime}
\newcommand{\hs}{\hspace{1mm}}
\newcommand{\nar}{New Astronomy Reviews}
\def\gsim{~\rlap{$>$}{\lower 1.0ex\hbox{$\sim$}}}
\def\lsim{~\rlap{$<$}{\lower 1.0ex\hbox{$\sim$}}}
\def\Msun {\,\mathrm{M}_\odot}
\def\Jcrit {J_\mathrm{crit}}
\newcommand{\rsun}{R_{\odot}}
\newcommand{\mbh}{M_{\rm BH}}
\newcommand{\Msunyr}{M_\odot~{\rm yr}^{-1}}
\newcommand{\mdot}{\dot{M}_*}
\newcommand{\ledd}{L_{\rm Edd}}
\newcommand{\cmc}{{\rm cm}^{-3}}
\def\gsim{~\rlap{$>$}{\lower 1.0ex\hbox{$\sim$}}}
\def\lsim{~\rlap{$<$}{\lower 1.0ex\hbox{$\sim$}}}
\def\Msun {\,\mathrm{M}_\odot}
\def\Jcrit {J_\mathrm{crit}}

\def\simgreat{\lower2pt\hbox{$\buildrel {\scriptstyle >}
   \over {\scriptstyle\sim}$}}
\def\simless{\lower2pt\hbox{$\buildrel {\scriptstyle <}
   \over {\scriptstyle\sim}$}}
\def\msobh{M_\bullet^{\rm sBH}}
\def\zodot{\,{\rm Z}_\odot}
\newcommand{\lambdabar}{\mbox{\makebox[-0.5ex][l]{$\lambda$} \raisebox{0.7ex}[0pt][0pt]{--}}}

\def\na{NewA}%
\def\aj{AJ}%
\def\araa{ARA\&A}%
\def\apj{ApJ}%
\def\apjl{ApJ}%
\def\jcap{JCAP}

\def\pasa{PASA}

\def\apjs{ApJS}%
\def\ao{Appl.~Opt.}%
\def\apss{Ap\&SS}%
\def\aap{A\&A}%
\def\aapr{A\&A~Rev.}%
\def\aaps{A\&AS}%
\def\azh{AZh}%
\def\baas{BAAS}%
\def\jrasc{JRASC}%
\def\memras{MmRAS}%
\def\mnras{MNRAS}%
\def\pra{Phys.~Rev.~A}%
\def\prb{Phys.~Rev.~B}%
\def\prc{Phys.~Rev.~C}%
\def\prd{Phys.~Rev.~D}%
\def\pre{Phys.~Rev.~E}%
\def\prl{Phys.~Rev.~Lett.}%
\def\pasp{PASP}%
\def\pasj{PASJ}%
\def\qjras{QJRAS}%
\def\skytel{S\&T}%
\def\solphys{Sol.~Phys.}%
\def\sovast{Soviet~Ast.}%
\def\ssr{Space~Sci.~Rev.}%
\def\zap{ZAp}%
\def\nat{Nature}%
\def\iaucirc{IAU~Circ.}%
\def\aplett{Astrophys.~Lett.}%
\def\apspr{Astrophys.~Space~Phys.~Res.}%
\def\bain{Bull.~Astron.~Inst.~Netherlands}%
\def\fcp{Fund.~Cosmic~Phys.}%
\def\gca{Geochim.~Cosmochim.~Acta}%
\def\grl{Geophys.~Res.~Lett.}%
\def\jcp{J.~Chem.~Phys.}%
\def\jgr{J.~Geophys.~Res.}%
\def\jqsrt{J.~Quant.~Spec.~Radiat.~Transf.}%
\def\memsai{Mem.~Soc.~Astron.~Italiana}%
\def\nphysa{Nucl.~Phys.~A}%

\def\physrep{Phys.~Rep.}%
\def\physscr{Phys.~Scr}%
\def\planss{Planet.~Space~Sci.}%
\def\procspie{Proc.~SPIE}%

\newcommand{\rmp}{Rev. Mod. Phys.}
\newcommand{\ijmpd}{Int. J. Mod. Phys. D}
\newcommand{\sovjetp}{Soviet J. Exp. Theor. Phys.}
\newcommand{\jkas}{J. Korean. Ast. Soc.}
\newcommand{\PPVI}{Protostars and Planets VI}
\newcommand{\njp}{New J. Phys.}
\newcommand{\rap}{Res. Astro. Astrophys.}

\title{Formation of the First Black Holes}

\setcounter{chapter}{10}

\chapter[Super-Eddington accretion]{Super-Eddington accretion; flow regimes and conditions in high-$z$ galaxies}\footnotetext{\tiny Preprint~of~a~review volume chapter to be published in Latif, M., \& Schleicher, D.~R.~G., ``Super-Eddington accretion; flow regimes and conditions in high-$z$ galaxies'', Formation of the First Black Holes, 2018 \textcopyright Copyright World Scientific Publishing Company, www.worldscientific.com/worldscibooks/10.1142/10652\par}\label{ch11}

\vspace{-30pt}

\author[Lucio Mayer]{Lucio Mayer}

\address{Center for Theoretical Astrophysics and Cosmology\\Institute for Computational Science\\University of Zurich\\
Winterthurerstrasse 190, CH-8057 Z{\"u}rich, Switzerland\\
lmayer@physik.uzh.ch}

\begin{abstract}
We review and discuss theoretical studies addressing the possibility of gas accretion onto black holes occurring 
at rates exceeding the Eddington limit. Our focus is on the applications to the growth of black hole 
seeds at high redshift. We first present the general notion of Super-Eddington accretion,
and then summarize the different 
models and numerical simulations developed to study such regime. We consider
optically thick flows in accretion disks as well as in spherically symmetric envelopes, and devote particular attention
to the widely adopted model based on the SLIM disk solution. While attractive for its simplicity,
the SLIM disk solution is challenged by the latest generation of three-dimensional radiation (magneto)-hydrodynamical
simulations, in which radiative losses can be an order of magnitude higher, and the mechanisms of radiation transport 
is more complex than straight advection as it takes place in a complex turbulent regime.
We then discuss the
gas supply rate to the sub-pc scale accretion disk or envelope from larger scales, revisiting gas inflow rates in protogalaxies
under various conditions. We conclude that in the dense gaseous nuclei of
high-z galaxies the conditions  necessary for the onset of Super 
Eddington  accretion regimes, such as a high optical depth and high gas supply rates from large scales,
should be naturally met. Feedback from the growing
BH seed should not alter significantly such conditions according to the results of radiation magneto-hydrodynamical simulations
of super-critical flows in accretion disks. Furthermore, based on the required nuclear gas inflow rates and the
tendency of stellar feedback to remove efficiently gas in low mass halos,  we argue that super-critical accretion will be more easily 
achieved in relatively sizable
halos, with virial masses $M_{vir} > 10^{10} M_{\odot}$, which become more common at $z < 15$.

\setcounter{page}{195}

\body

\end{abstract}

\section{Introduction}

In this chapter we will cover and summarize work carried out to model
Super-Eddington flows on massive black hole seeds. These are flows in
which the accretion rate onto a massive seed black hole occurs at a rate
higher than predicted by the Eddington limit. 
We refer to Chapter 10 for the basic concepts of accretion onto
black holes and the general understanding of massive black hole growth
on cosmic timescales. Here we recall that for a black hole of mass $M_{BH}$, the
Eddington limit on the mass accretion rate $\dot M_{Edd}$ is set by imposing perfect balance
between the gravitational force pulling matter towards the black hole and
radiation pressure of the emerging photons exerting a force in the direction
opposite to the accretion flow. It is obtained under the assumption of
spherical symmetry and is given by $\dot M_{Edd} = 4 \pi G M_{BH} m_p/\epsilon c \sigma_T$,
where $G$ is the gravitational constant, $m_p$ is the proton mass, $\epsilon$ is the radiative
efficiency of the accretion process, and $\sigma_T$ is the Thomson scattering cross
section for the emerging photons by which radiation pressure is communicated to matter.
We also recall that the characteristic accretion timescale expressed as a function
of the Eddington luminosity $L_{Edd}=  \epsilon \dot M_{Edd} c^2$ reads:

\begin{equation}
t_{\rm acc}=\left({\epsilon\over 1-\epsilon}\right)\left({L_{Edd}\over L}\right)t_E=(4.3\times 10^7 {\rm 
yr})\,\left({L_{Edd}\over 
L}\right),
\end{equation}
where $t_E = M_{BH}c^2/L_{Edd}$ is the Eddington timescale, and
the last equality assumes $\mu_e=1.15$ (valid for primordial gas) and a radiative efficiency of $\epsilon=0.1$.

As mentioned in Chapter 10, unless  the formation of massive black hole seeds by
e.g. direct collapse is invoked, Super-Eddington accretion phases are required in order to grow
light seeds from Pop. III stars, with typical masses 
$\sim 100 M_{\odot}$ or lower,  to the gargantuan masses of  bright QSOs, $10^9-10^{10}  
M_{\odot}$ \citep{Mortlock2011,Banados18} by $z=6$.
It is thus of crucial interest to understand if Super-Eddington accretion
can occur under realistic conditions.
 
\begin{figure}[h]
\includegraphics[width=34pc]{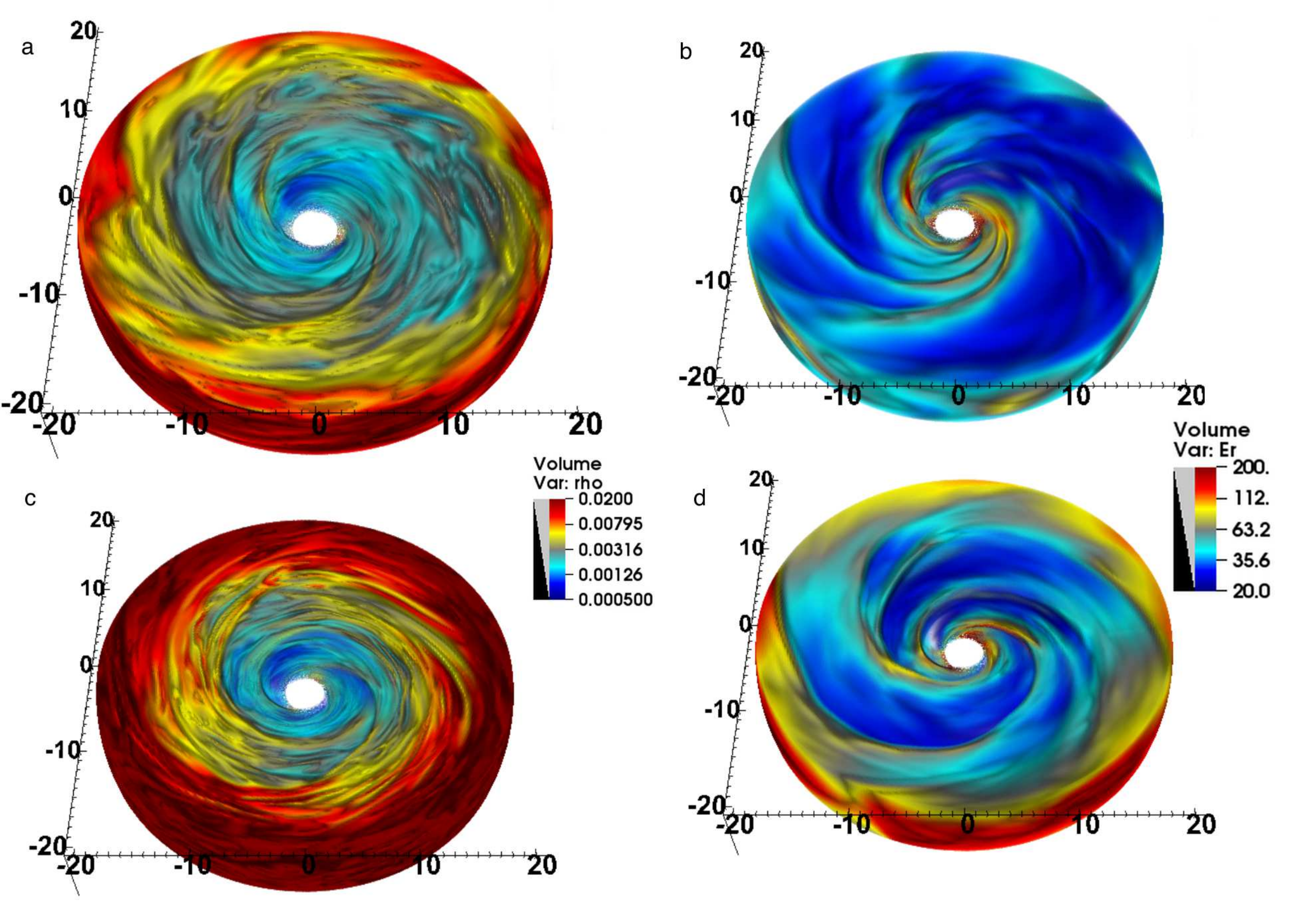}\hspace{2pc}
\caption{\label{label} Density (left panels) and radiative flux (right panels) snapshots from the 
\citet{Jiang17} 3D MHD simulations.
carried out with the VET method for radiative transfer. The highly turbulent nature of the flow, the generation of large scale 
spirals and the fact that the radiative flux has marked local variations are all features evident from the color-coded maps(see text).}
\end{figure}

For astrophysically relevant models of Super-Eddington accretion
there at least two key aspects 
that warrant consideration. One is, of course, under which conditions an accretion disk
or, more generically, a gaseous envelope, can supply gas to a black hole seed at a rate higher than
Eddington, and whether this can be achieved in a nearly steady-state regime
or only episodically  \citep{Pezzulli16}.
The other aspect concerns understanding the conditions under which
the accretion disk/envelope can be fed at a rate
well above Eddington via gas inflows initiated at much larger scales.
Tackling both requires studying and modeling the gas flow along with the various 
physical processes governing its behaviour across a prohibitively wide range of scales,
from several kiloparsecs to centi-parsecs. Such complexity is conceptually and computationally
challenging, and is a absent in the case of stellar mass black holes
for which  only the scale of the accretion disk is
relevant as external gas flows are negligible (unless the black hole is
part of a binary system, in which case other considerations may apply).
The problem of studying gas inflows at galactic scales is 
already by itself a multi-scale process since its natural boundary conditions
are determined by gas dynamics in the circumgalactic medium. This is especially true when
a galaxy is still in the early assembly  phase and thus is hosted 
in a dark matter halo vigorously fed by
cold filaments originating from the cosmic web (the so-called "cold accretion"
mode, see e.g.  \citet{DekelBirnboin2006,Keres2009}.
The timescales of gas inflows may vary from scale to scale as different phenomena
drive them, e.g. nuclear bars and spiral density waves in galactic nuclei, tidal
interactions, mergers and bar-instabilities on galactic scales, cold flows and feedback-driven
galactic fountains on larger scales. Ultimately, the duty cycle of gas feeding
to the black hole seed will be determined by the concerted action of all these
processes, hence a characteristic timescale is hard to define. Because of that
the simplest approach will have to assume a boundary value for $\dot M$, constant
or time-dependent, and explore whether or not Super-Eddington regimes can persist.
This is indeed the approach followed by numerical simulations that model directly
a accretion in a disk/envelope surrounding the black hole.

While the possibility of super-critical accretion onto black holes 
has been considered since a long time
\citep{Begelman79,Abramowicz88}, it is only recently that hydrodynamical simulations have become
capable of studying its feasibility in realistic astrophysical environments, including in the
specific case targeted by this chapter, namely  massive black holes  and black hole  seeds.
Pioneering numerical work over a decade ago has been carried out by
\citet{Ohsuga2005}, who identified very clearly two key factors leading to super-critical
accretion; photon trapping and the anisotropy of the radiation field. These two aspects play a
different role in the various super-critical simulations of accretion disks or accreting spherical envelopes
that were developed later, as we will discuss in this chapter.

Following the rationale of the book we will focus on  host galaxies 
at very high redshift, between $z=20$ and $z=6-7$, namely up to as late as
the epoch of appearance of the first QSOs (as early as $z=7.5$ according to \citet{Banados18}.
We recall that the first motivation for studying Super Eddington accretion at such
early epochs is to find out if light BH seeds formed by the collapse of primordial metal-free
Pop. III stars could grow rapidly and become as massive as required to explain
the population of bright high-z QSOs at $z > 6$ discovered in the last decade (see
Chapter 12).
It is indeed the only plausible alternative route
to postulating a direct collapse scenario, a topic covered in other chapters of this book.
As we will see, though,  simulations are revealing
that galaxy assembly is highly dynamical and complex at such high redshifts,
and has wide range of flow conditions. Therefore, the nature and strength of gas inflow rates at scales 
above the accretion disk/envelope depend on galaxy mass as well as on the nature of
gasdynamics in the host galaxy.
In other words, galaxies at such high redshifts are by no means simple. Observational
evidence on the nature of gas flows and stellar components of galaxies is still lacking, so this is still mostly
theorists' playground. Yet the advent of ALMA, JWST, WFIRST, EUCLID and large ground based telescopes
such as E-ELT will change the landscape soon, not only by providing enough sensitivity
to detect emission from the early stages of massive BH seed growth, but also allowing to 
distinguish, in principle, between different scenarios for their initial birth and subsequent growth,
including  eventually the regime of super-critical growth (see chapter 14).

The chapter is organized as follows. In section 2 we begin by discussing theoretical scenarios for 
the realization of Super-Eddington accretion, starting from relatively idealized models
such as the popular SLIM disk model, and we will then move on to discuss the results of complex
three-dimensional simulations that include radiation, the effect of the magnetic and
, in some cases, also solve the set of equations in the relativistic regime. A major distinction,
in this context should be done between accretion occurring in a rotationally supported configuration,
namely an accretion disk, and accretion occurring in a spherically symmetric envelope. For the latter reason
two separate subsections are devoted to the two different flow regimes. Section 3 concerns the conditions
of the large scale gas inflows in protogalaxies and galaxies at high-z, namely focusing on how we
can inform the boundary condition for the accretion flow onto a black hole seed in a galactic/proto-galactic nucleus.
Finally, section 4 describes the first attempts made to combine models of super-critical accretion such
as those discussed in section 2 with the galactic astrophysics framework covered in section 3. 
hydrodynamical simulations of galactic/protogalactic nuclei. We conclude with section 5, which summarizes
the whole chapter.

\begin{figure*} 
\includegraphics[width=5.0 cm]{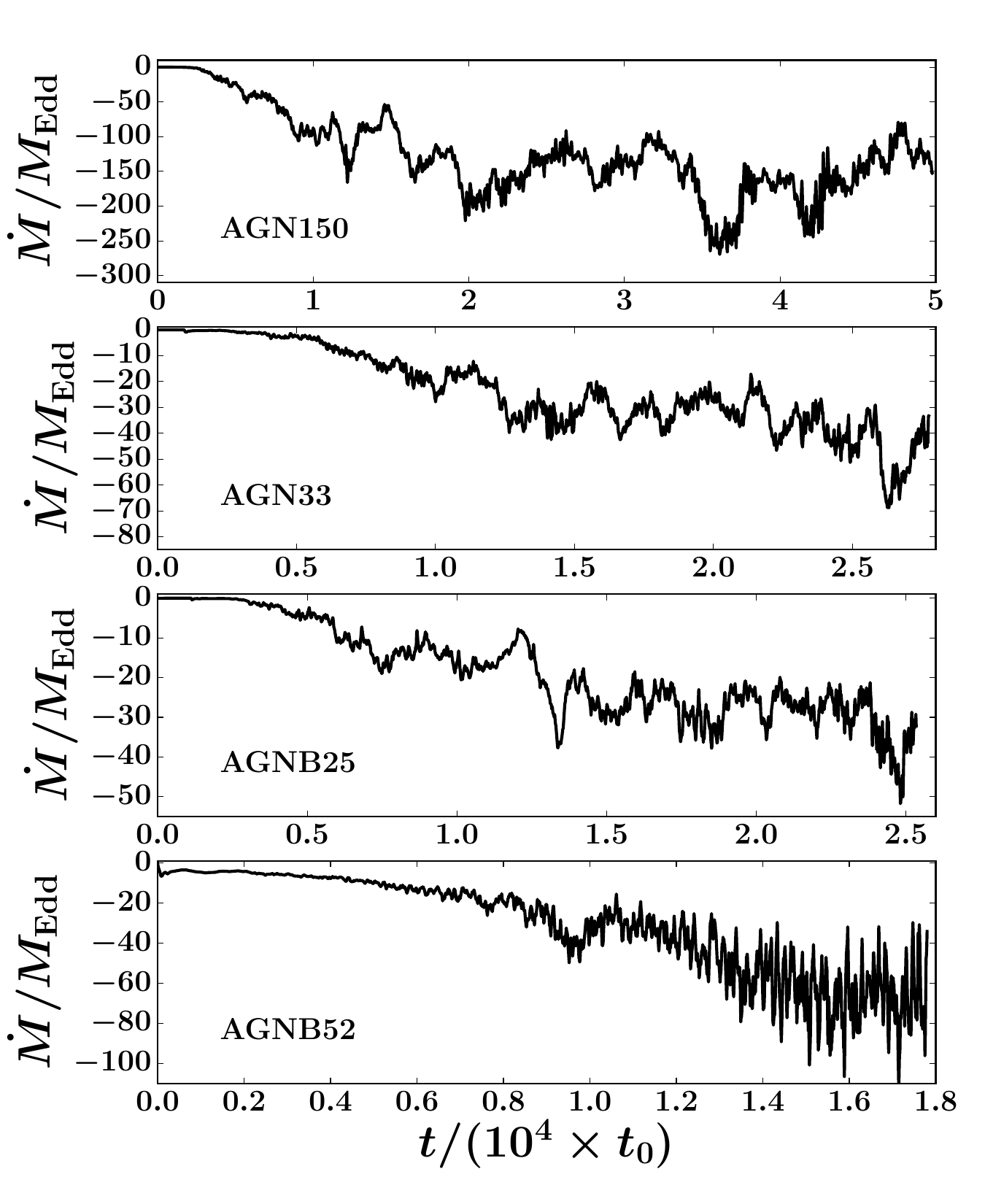} \hspace{0.2 cm}
\includegraphics[width=5.0 cm]{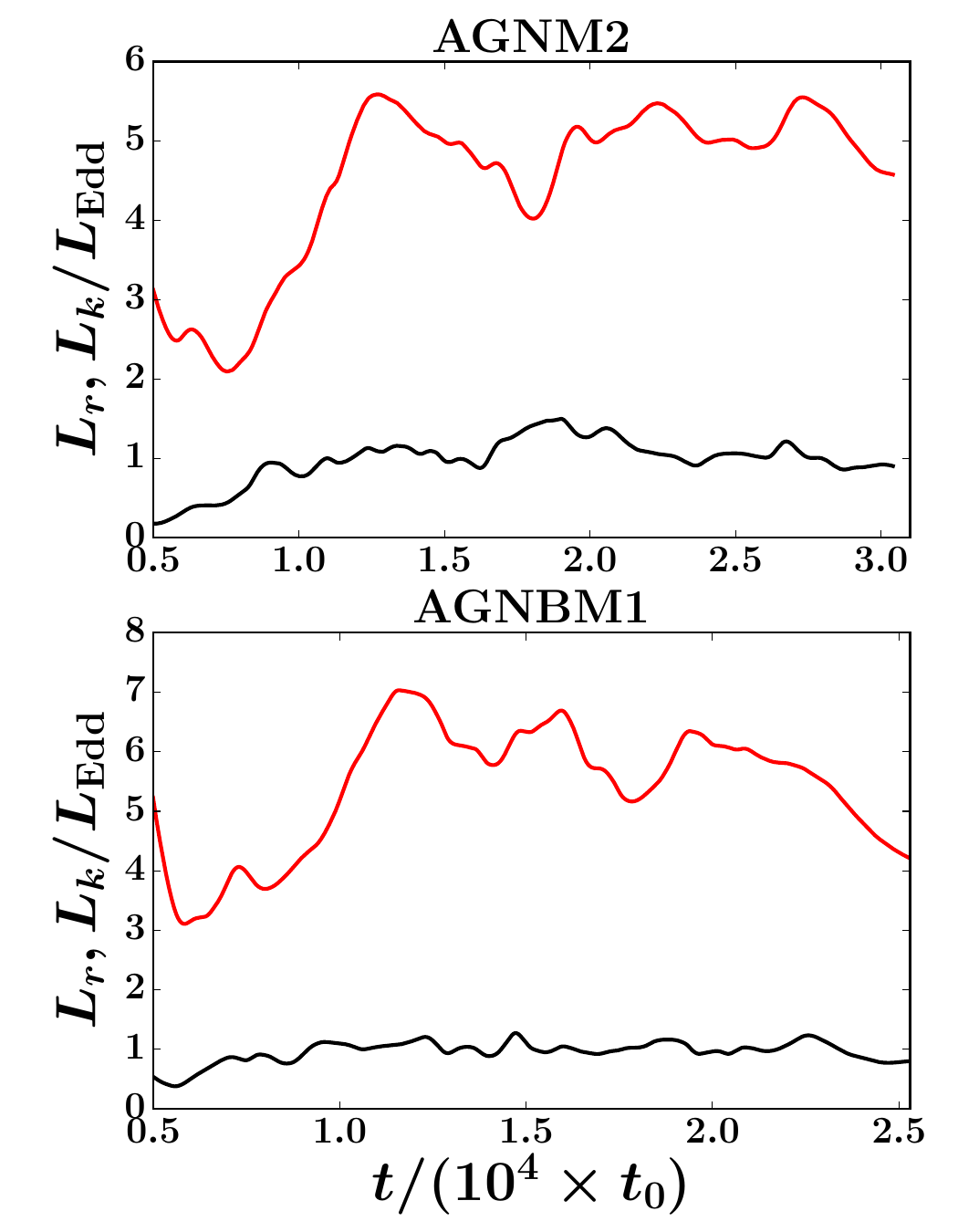}
\caption{\label{label} In the left panel we show the history of the black hole accretion rate at 10 Schwarzschild radii
for four representative simulations of accreting SMBHs  in \citet{Jiang17}, in units of the accretion 
rate corresponding
to the Eddington limit. In the left panel the radiative and
kinetic luminosity in units of the Eddington luminosity, time-averaged and plotted as a function of distance,
are shown for two of such runs. Time is shown in a dimensional units, the same in each run.}
\end{figure*}

\section{Super-Edddington flows: direct models of the accretion process}

This section is devoted to discuss theoretical models and simulations carried out at scales
smaller than the Bondi radius of the black hole to study directly the accretion process,
eventually down to the radius of the last stable orbit. The works summarized here represent
the backbone of our current understanding of Super-Eddington accretion. There is, however, no direct
information on nuclear and galactic-scale gas dynamics as these models, due to their
computational complexity, cannot cover a wide range of scales. We have divided the section into two main
parts, one focusing on flows in accretion disks and one on flows in extended envelopes that are not
rotationally supported.

\subsection{Super-Eddington flows in accretion disks: from the SLIM disk models to 3D MHD simulations}

For accretion flows occurring in rotating disk-like configurations 
the SLIM disk model  \citep{Begelman79,Abramowicz88} has is a well-established framework yielding
well understood super-critical flow solutions.
Analogously to Shakura-Sunyaev "alpha" disks, in SLIM disks dissipation is
described by a viscous stress tensor that is proportional to pressure.
Slim disk solutions are obtained by solving vertically-averaged radial momentum
equations. Later \citet{Sadowski11} improved the model introducing
a vertical structure by solving the the fully relativistic, axisymmetric hydrodynamical equations
in a Kerr metric augmented with hydrostatic equilibrium, a vertical energy profile and
energy transport differential equations.
In this steady-state model accretion can occur in a Super-Eddington regime if the medium
is optically thick, so that radiation is
advected with the flow rather than escaping. In order for the latter condition
to be satisfied the photon diffusion timescale has to be much longer than the
radial advection timescale, which is governed by viscous transport. By construction
this requires a very high optical depth, and indeed steady-state models such as
those in \citet{Sadowski11} have an optical depth $\tau > 10^3$ beyond ISCO
(the innermost stable circular orbit around a Kerr black hole).
Note that, because of the high optical depth, the vertical energy transport 
equations can be solved using the diffusion approximation \citep[as in][]{Sadowski11}.
The timescale constraints in the SLIM disk model imply that no outflows are possible
because, for that to happen, radiation pressure has to overcome bulk thermal and turbulent
pressure. This result is at variance with what will happen in a standard radiative accretion disk
in which, as the inflow rate rises above the Eddington limit, a radiation 
pressure-driven outflow would occur. As we will see later, realistic 3D MHD
simulations suggest that neither picture is correct, namely while
advection is important and partial photon trapping can occur in a dense, heavily
mass loaded accretion disk, outflows can also occur. The extent by which outflows
are important, and how luminous above Eddington can the disk be, is dependent on the nature of radiative transport.
The jury is still out on which numerical approach to radiative transport is capturing the energy flow via radiation
most realistically. Therefore, this is an area of research in which new insight is strongly tied to advances in
simulation methods.

\begin{figure}[h]
\includegraphics[width=26pc]{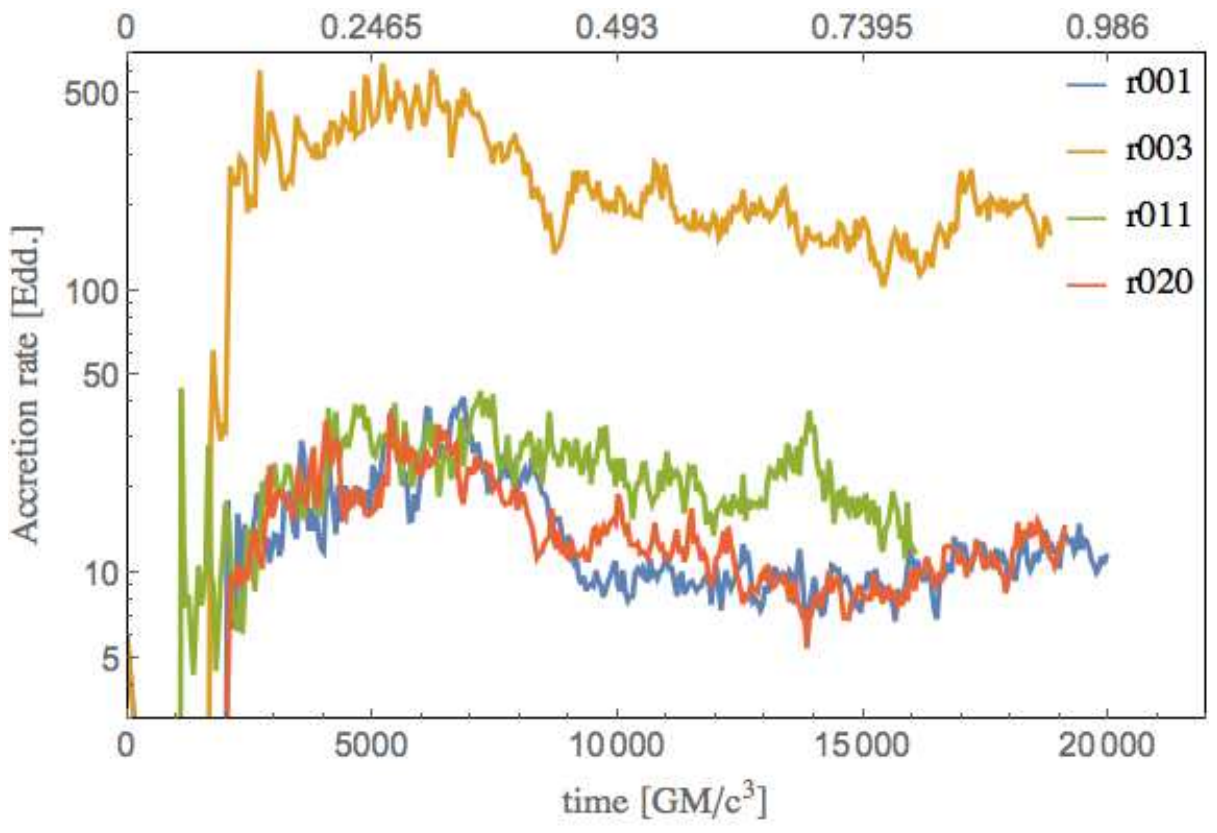}\hspace{1pc}
\caption{\label{label} Accretion rate histories in units of the Eddington accretion rate for four of the
3D relativistic runs of \citet{Sadowski16} with stellar mass black holes (see text). Figure adopted from \citet{Sadowski16}, reproduced by permission of Oxford University Press / on behalf of the RAS.}
\end{figure}

A straightforward way to characterize the SLIM disk
model is to state that accretion is highly radiatively inefficient compared to the standard
efficiency in Shakura-Sunyaev accretion disks around a Kerr black hole. This is because
the  photon diffusion timescale becomes much longer than the other characteristic timescales
in the disk, namely the sound crossing timescale and the viscous timescales associated with
the mechanism of mass transport.
Indeed SLIM disk
solutions imply a radiative efficiencies low as low as $< 0.1\%$ \citep{Sadowski13,Sadowski16}
as opposed to $\sim 40 \%$ in a standard viscous disk around a spinning BH.
As a result, $\dot M$ can be 1-2 orders of magnitude
above $M_{EDD}$ while the emitted luminosity remains close to the Eddington limit.
This has important consequences on the energetic
feedback onto the surrounding interstellar medium, as discussed below (see
section 4). A  low radiative efficiency also characterizes the advection dominated flow (ADAF) model,
except that in that the medium is optically and geometrically thin, which 
can only occur with very low accretion
rates  (Sub-Eddington). 
The latter is the opposite situation of what is needed
in the SLIM disk model, which requires a medium  that is optically
thick, which yields long diffusion timescales, and thus relatively hot.
An attractive feature of the SLIM disk model is that
a simple fitting formula for the radiative efficiency, weakly dependent
on the spin of the Kerr black hole, can be found \citep[see][]{Sadowski9,Madau2014}. The latter
was readily employed
in large-scale MBH accretion models in place of the standard viscous disk radiative
efficiency, as we will discuss in section 4.

\begin{figure*}
\includegraphics[width=5.4 cm]{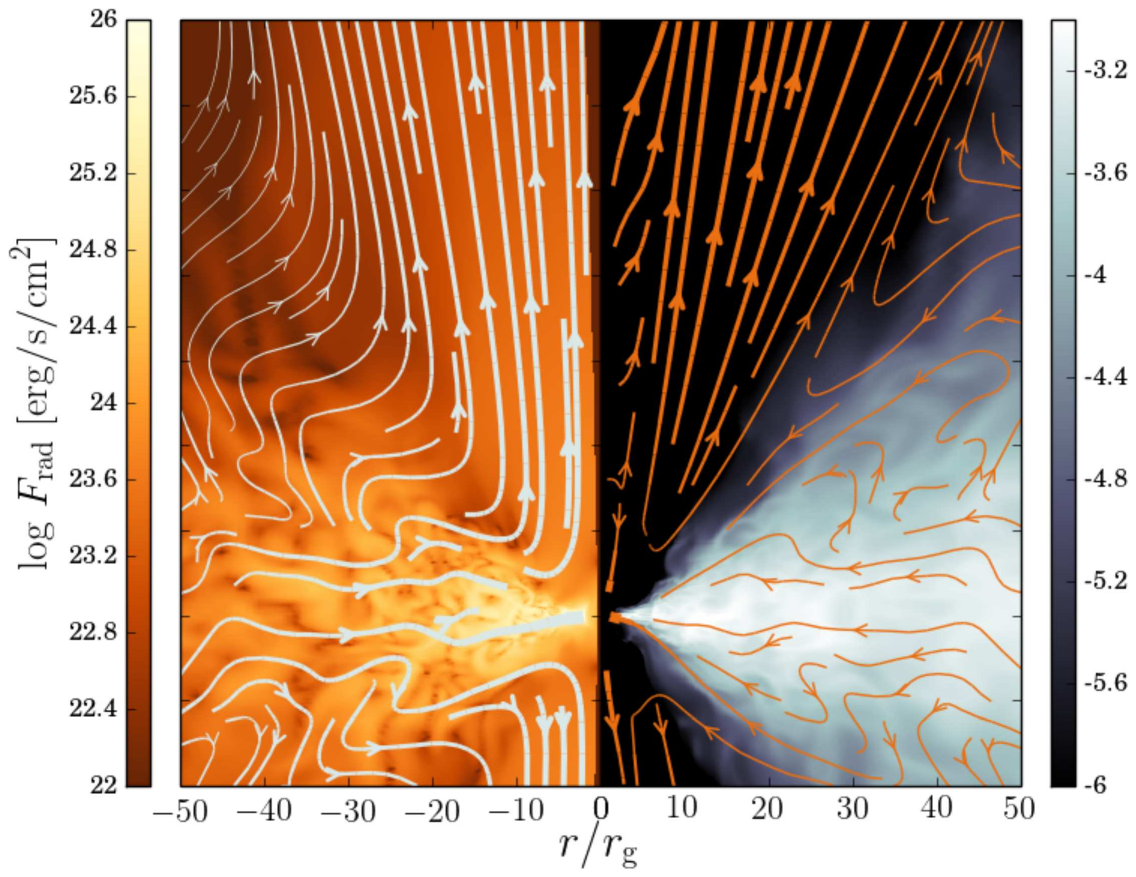} \hspace{0.2 cm}
\includegraphics[width=5.4 cm]{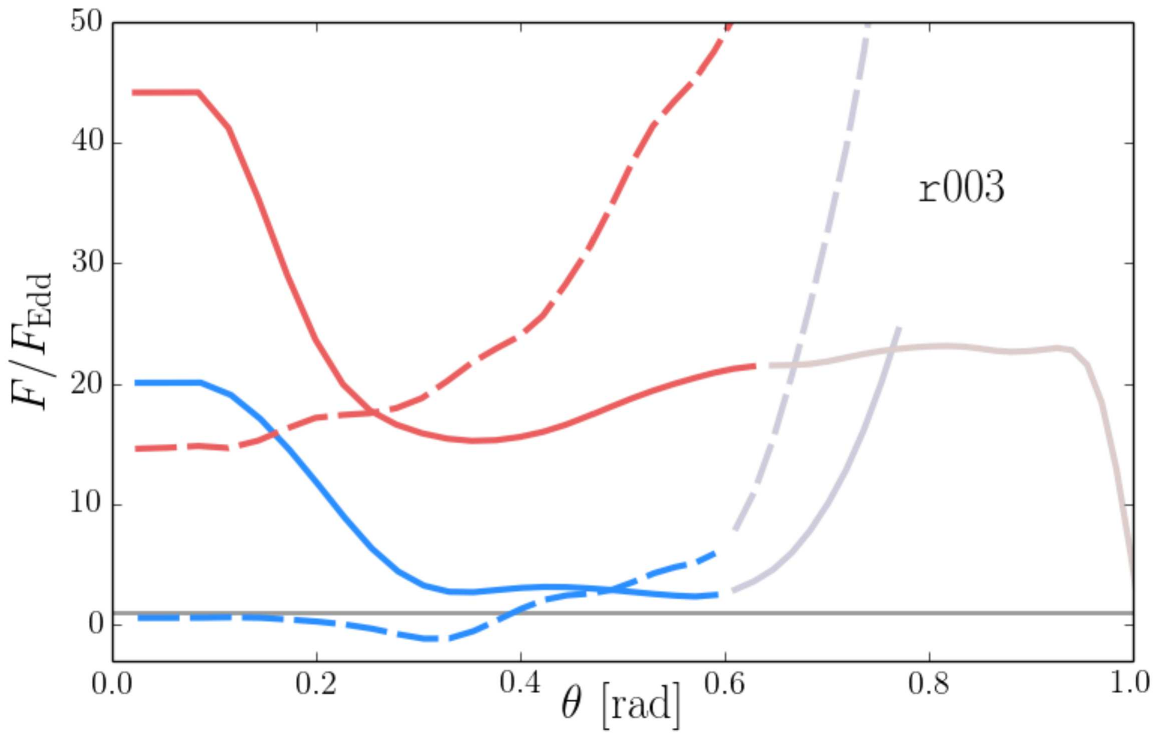}
\caption{\label{label} On the left a slice through the midplane of the accretion disk at the end
of one of the simulations of \citet{Sadowski16} shows the radiative flux (orange-scale map) and the 
density map (grey-scale map).The powerful winds leaking out radiation via advection are evident, as is the photon
trapping through inward advection in the high density mid-plane. On the right we show an example
of the radiative flux (solid lines) and kinetic flux (dashed lines) in units of the  Eddington flux from one of the runs, 
using  blue for quantities measured near the edge of the disks and red for quantities
measured near the inner annulus of the computational domain. From the plot
it is evident that the kinetic luminosity can supersede the radiative luminosity, in contrast
with \citet{Jiang17} (see text).}
\end{figure*}

While 2D radiation-hydro simulations of accretion disks or spherically symmetric
envelopes have been available since the early 2000s (see \citet{Ohsuga2005} and section 1),
it is only recently that simulations that include the third dimension as well as MHD
to study the Super-Eddington regime have appeared in the literature (see example in Figure 1,
from \citet{Jiang17}.
Currently there a just a handful
of 3D super-critical accretion disk simulations. They are carried out by different
research groups employing different numerical techniques and approaches,
each with its own advantages and disadvantages.  Overall they can achieve $\dot M$ in the range several tens
to a few hundred times $\dot M_{Edd}$ (Figure 2 and 3).
A large, super-critical $\dot M$ is
imposed as a boundary inflow condition at the beginning of the simulation \citep{Sadowski16,Jiang17}.
Modern MHD simulations
are carried out at high resolution with low-diffusivity numerical schemes that
can capture accurately the magneto-rotational instability (MRI). The latter is
found to be the main 
angular momentum and mass transport mechanism in the accretion disk, in agreement
with the findings for the more conventional regime of accretion in presence of low rates
of mass loading. MRI is thus the main source of viscous transport,
although evidence is mounting that in dense optically thick disks other transport mechanisms
are also at play.

One group of simulations, primarily by Sadowski
and collaborators, comprises {\it fully relativistic} 3D MHD calculations with a finite
difference code combined with a relatively simple treatment of the radiation part, either
using flux-limited diffusion \citep[FLD, e.g.][]{Sadowski13}, or, more recently, the M1 approximation \citep{Sadowski16}. These simulations so far have focused on accretion onto stellar
mass black hole rather  SMBHs, although there is one single attempt to study the accretion
around a 1000 $M_{\odot}$ MBH in \citet{Sadowski16} which yielded qualitatively
similar results.

\begin{figure*}[h]
\includegraphics[width=5.5 cm]{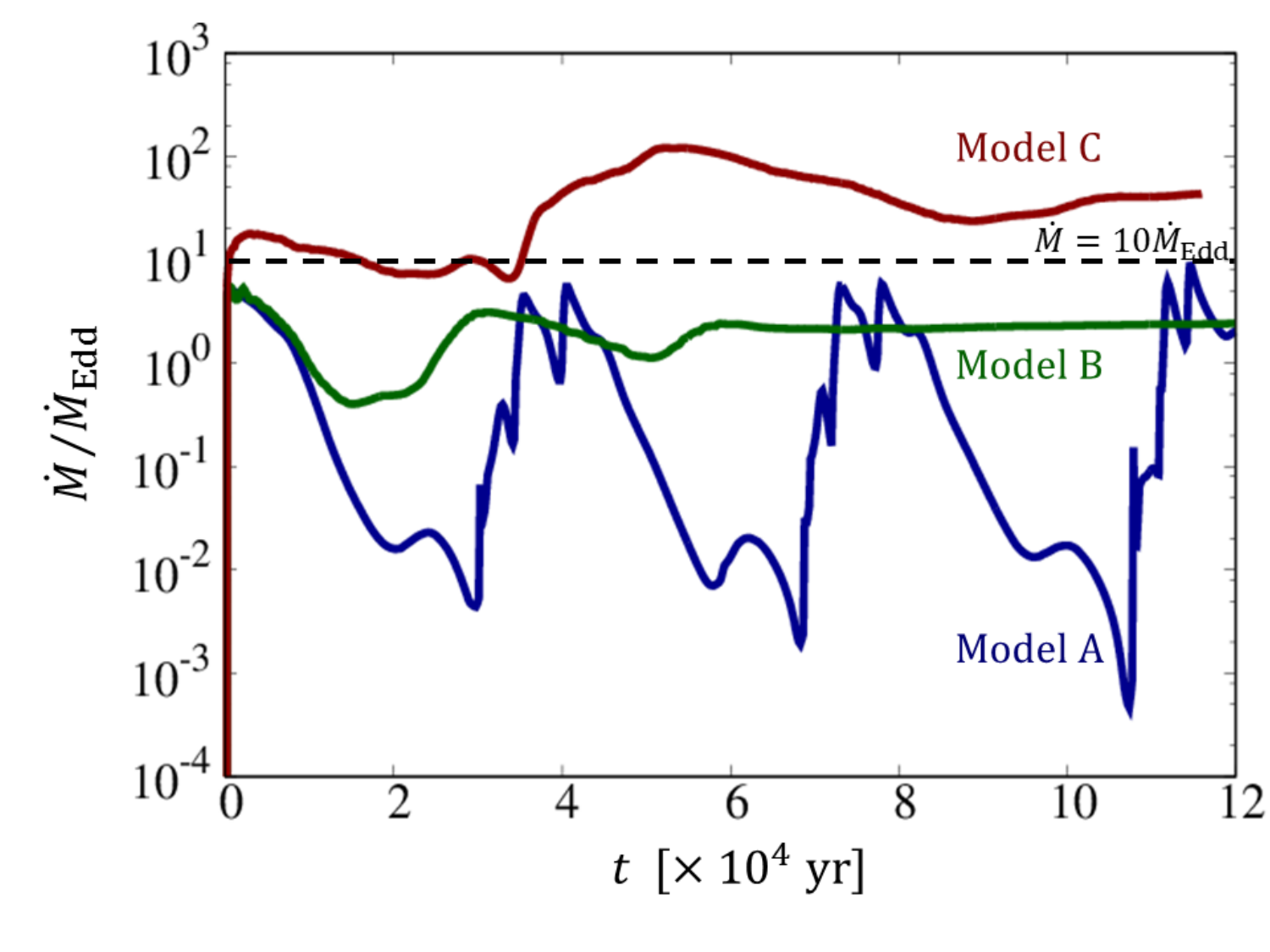}
\includegraphics[width=5.5 cm]{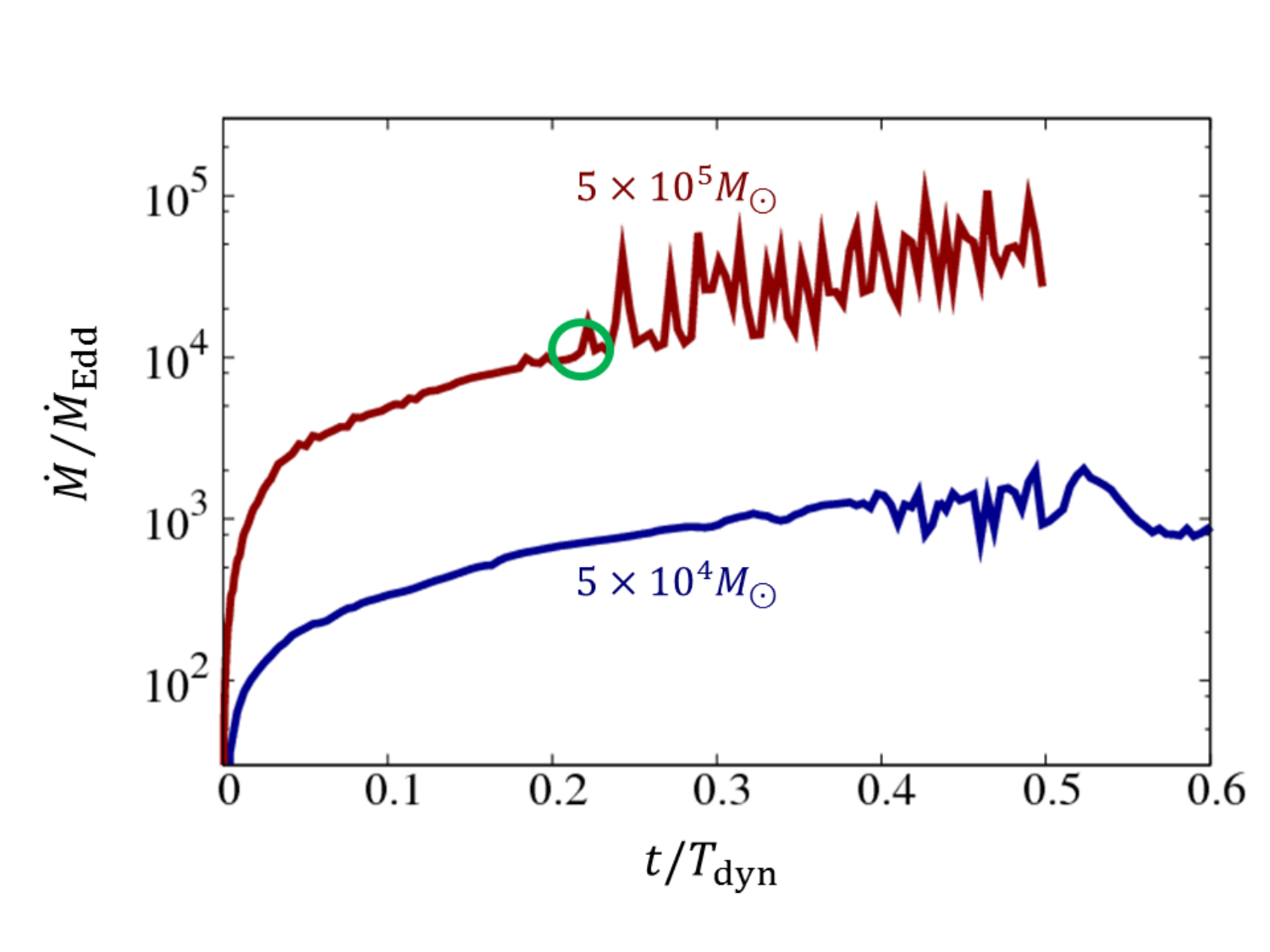}
\caption{\label{label} Examples of Hyper-Eddington accreting light (left) and massive BH seeds
(right) from  \citet{Takeo18}. The mass of the BH is $10^3 M_{\odot}$ on the left, and the
different models differ for how the radiation field geometry is imposed (isotropic for Model A and increasingly
aniostropic for Model B and C). Super-critical accretion sets in for anisotropic radiation propagation.
For seeds with masses $> 10^5 M_{\odot}$, shown on the right, the ionization region can recombine rapidly and collapse,
leading to the highest accretion rates (after the time marked with the green circle). Figure adopted from \citet{Takeo18}, reproduced by permission of Oxford University Press / on behalf of the RAS.}
\end{figure*}

A second type of simulations \citep[Jiang et al. 2014, ][]{Jiang17}
comprises {\it non-relativistic} 3D MHD calculations with finite volume hydro methods (ATHENA) using a 
highly sophisticated
radiative transport schemes such as the variable Eddington tensor (VET) scheme coupled
with solving the steady-state radiative transfer equation at each timestep. With
the latter approach the very inner region of the disk, including the generation of a relativistic
jet, cannot be properly modeled,  but radiation transport is more accurate and realistic as 
it takes into account the actual flow configurations. This can have significant effects 
on the nature of energy and momentum transport through the disk as well as on the net energy loss. 
Very recently the latter simulation method has been applied to studying accretion onto a SMBH of mass
$10^8 M_{\odot}$ \citep{Jiang17}.

Perhaps not surprisingly given the differences in the numerical methods, the results of simulations
carried out by different groups are quite different. Both the radiative efficiency and the
relative proportion between the radiative and mechanical energy output generated by accretion
differ significantly.
In particular, the simulations with the VET method
produce higher radiative efficiencies, at least one order of magnitude above those of \citet{Sadowski16}.
The latter result is still mostly based on studying accretion onto stellar 
mass black holes. \citet{Jiang17} have suggested that the cause of the difference might be 
the more pronounced anisotropy of the radiative flux captured with the  VET 
method as opposed to the M1  method. The M1 approximation is indeed still a moment-based
approach to the radiative transfer equation, which means it still integrates out the information
on directionality of photon propagation to some extent.
A higher anisotropy in the propagation pattern  allows radiation 
to percolate more efficiently through a clumpy, turbulent medium with
large inhomogeneities. MRI, as expected, gives rise to a turbulent flow
regime, which naturally creates a highly inhomogeneous flow
(Figure 1).

When compared to the standard SLIM disk model, the simulations 
of \citet{Jiang17}, which are the only ones to focus exclusively on 
accretion onto supermassive black holes,
are  those that show the most striking differences. Photon
transport occurs via advection, not diffusion as in the SLIM disk model, and disks
evolve in a regime in which the advection timescale is shorter than the sound
crossing timescale. Advection is primarily generated by MRI turbulence. Advection
of photons along with the gas flow is also the main mechanism of radiative emission
in the simulations of Sadowski and collaborators. Despite higher
luminosities, because accretion and emitted radiation are both highly anisotropic, high
mass accretion rates exceeding 100 times the Eddington rate do occur (Figure 2).

In addition, \citet{Jiang17} show that radiation can leak out also via other forms of turbulent transport,
in which, while mass is not advected, radiation can be advected due to effects such as
magnetic buoyancy. {\it In this case radiation transport is not through diffusion nor through
advection, rather it occurs in a third way which is of course not captured by simple
models.} On the contrary, \citet{Sadowski16} found turbulent transport of radiation
to be negligible compared to advection in their simulations. The reason of the latter discrepant
conclusion is still unclear. While Sadowski and collaborators find a higher kinetic luminosity,
especially in the outer parts of the disks (Figure 4), the total luminosity in their most recent
three-dimensional simulations \citep{Sadowski16}, hence the sum of kinetic and
radiative luminosity, is found to be only a factor of 2 lower than that expected in the standard
viscous thin disk accretion theory for both zero and high spin BHs, being, respectively, 
$3\%$ and $9\% \dot M c^2$.  In  \citet{Jiang17} the kinetic energy luminosity is always $15\% - 30\%$
of the radiative luminosity, and increases with higher accretion rates..

Finally, while there is discrepancy in the predicted 
radiative luminosity,
both groups find radiative efficiencies higher than predicted by the SLIM disk model in the regime
of very high accretion rates, corresponding to at least 10 times the accretion rate at the 
Eddington limit.
For lower accretion rates the radiative efficiencies of \citet{Sadowski16}
agree with the predictions of the SLIM disk model, while those of \citet{Jiang17}  are in the
range $5-7\%$, hence significantly higher.
In  \citet{Jiang17} , the radiative efficiency first increases
with increasing accretion rate, then, as the accretion rate becomes higher than $1000 M_{\odot}$/yr,
it starts to decrease to $\sim 1\%$, a finding that is further supported by 
additional numerical simulations being analyzed as we write (Y. Jiang, private communication). 
 As a reference,
\citet{Sadowski16} find typically a radiative efficiency of $0.006$ and $0.01-0.02$ for, respectively, zero spin 
and near maximally rotating black holes. This is about an order of magnitude lower than 
the standard radiative efficiency in a thin viscous disk.

\begin{figure*}[ht]
\vspace{1.2cm}
\includegraphics[width=5.5 cm, height=5.5cm]{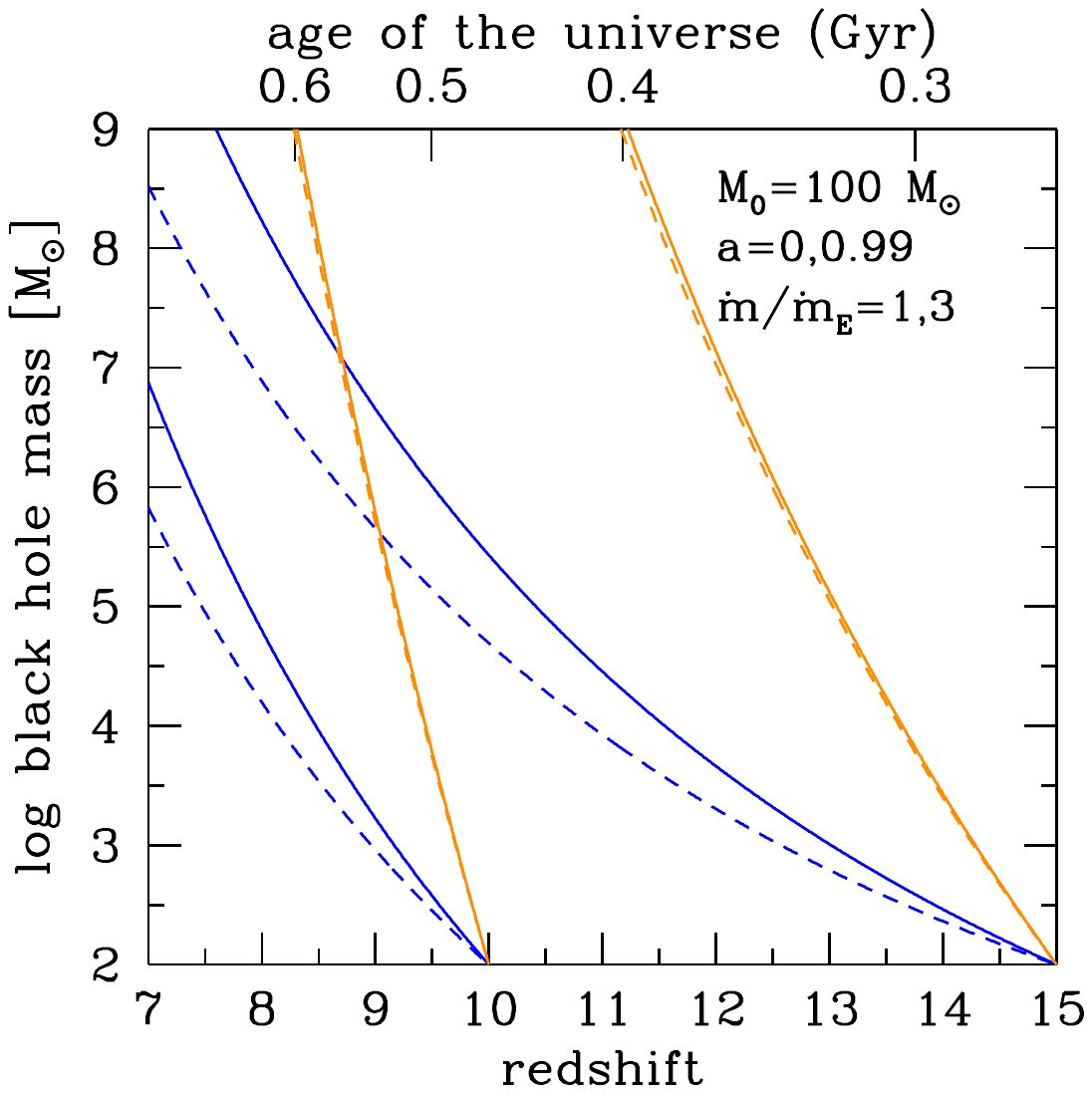}
\includegraphics[width=5.5 cm, height=5.5cm]{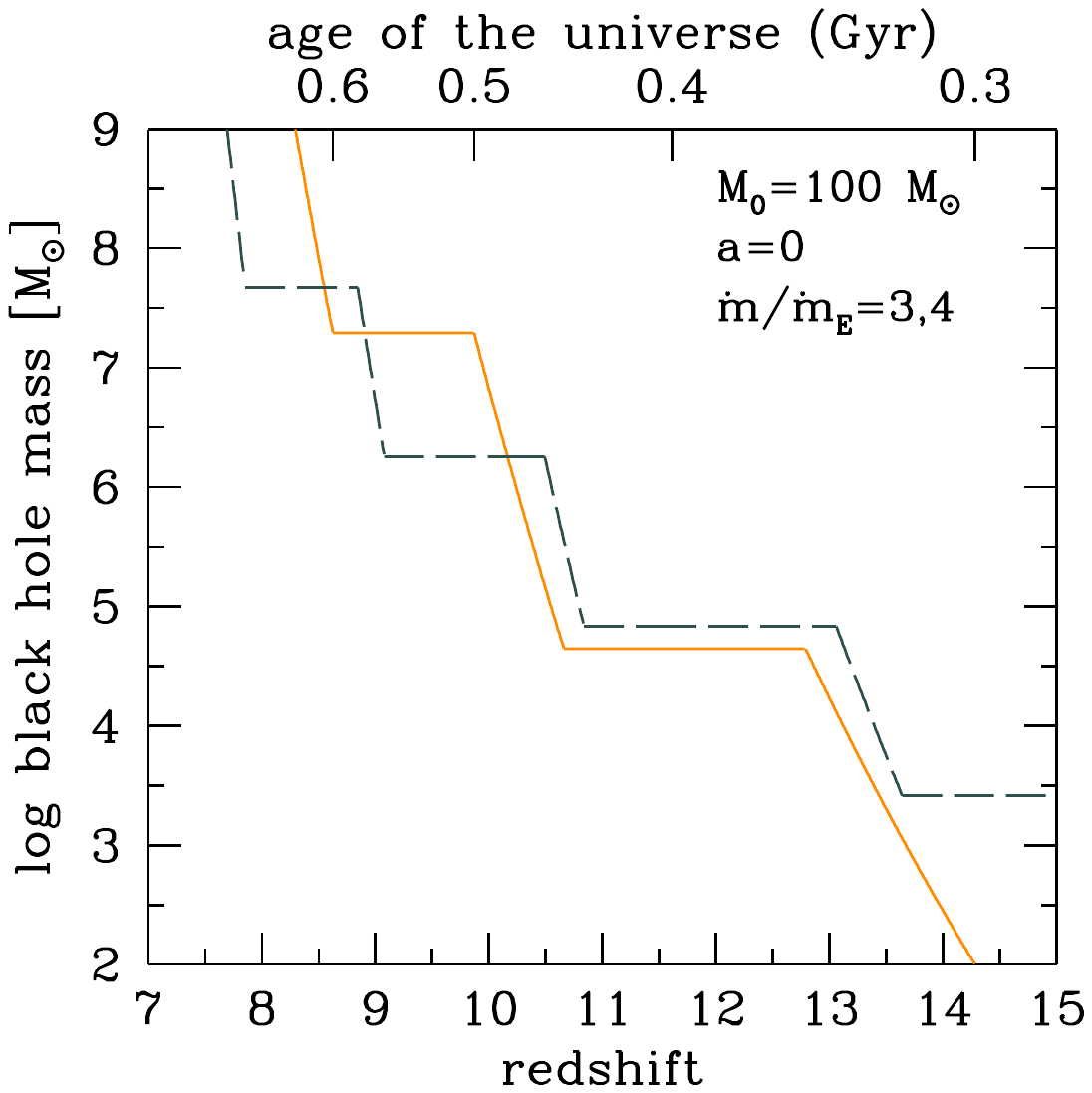}

\caption{\label{label} The left panel shows the different growth rate of an individual light BH seed,
with mass of $100 M_{\odot}$,
for models adopting, respectively, SLIM disk accretion (orange, for an accretion rate constantly at
3 times the Eddington rate) and a standard viscous accretion  at the Eddington limit (blue). Solid and dashed lines are
for, respectively, a zero spin and a highly rotating black hole (see labels in panel). The right panel shows
only super-critical growth curves for the same BH mass but for intermittent
accretion with a duty cycle of 0.5 (solid line, for 3 times the Eddington rate) and 0.2 (dashed line, for 4 times
the Eddington rate). Redrawn with permission from Madau et al. (2014).}
\end{figure*}

\subsection{Super-Eddington flows in spherically symmetric envelopes}

In this section we will discuss spherically symmetric accretion flows, without dynamically
important angular momentum. This
is the configuration in which the conventional Eddington limit can be applied more naturally,
as this assumes no directionality in the gas flow or in the radiation field. Nevertheless,
the Super-Eddington accretion regime is possible under certain conditions. 
Radiation hydrodynamics simulations have been used since more than a decade 
to show that, if the BH is embedded
in an optically thick medium, efficient photon trapping can reduce the radiative force
enough to allow super-critical accretion from an envelope  \citep{Ohsuga2005}. As in the case of accretion
disks, though, the problem has always been how to maintain a sustained super-critical regime.
The problem often documented in the literature is that, as soon as radiative feedback from the accreting
hole is accounted for, the emerging hot ionizing flux stifles gas accretion from larger scales
effectively  \citep{Milosavljevic09,Alvarez2009,Park:2011p780,Park12,Park:2013p2747}.
Therefore. the time-dependent infall rate at the edge of the envelope, a boundary
condition that in principle could be set based on gas inflow rates found at larger scales 
(e.g. in protogalaxy simulations), and the anisotropy of the radiation field, are both important
factors.  \citet{Inayoshi_Haiman_2016} have developed a self-consistent
model in which radiative heating and ionization is accounted for in 1D radiation-hydro
simulations. They found a steady-state solution corresponding to Super-Eddington accretion
at the remarkably high rate of more than  $5000 L_{edd}/c^2$, which they dub
as {\it Hyper-Eddington accretion} (Figure  5).
The configuration is that of a radiation-dominated core where photon trapping occurs surrounded by an accreting optically-thin
envelope. The core thus shields the envelope from the effect of radiation pressure.
The BH needs thus to be embedded in a dense gas cloud satisfying the following condition:

\begin{equation}
M_{\rm BH} \ge 10^4 M_{\odot}
\left(\frac{n_\infty}{10^5 cm^2}\right)^{-1}\left(\frac{T_\infty}{10^4 K}\right)^{3/2},
\end{equation}

where $n_\infty$ and $T_\infty$ are density and temperature of the ambient gas, respectively.
The condition above corresponds to the Bondi radius being larger than the size of the ionizing region
(the Stromgren sphere). If this is satisfied accretion occurs essentially at the Bondi rate. We recall that the Bondi accretion rate, which is
derived under the assumption of spherical symmetry and an infinitely large radius at which gas begins
to acquire a net inward radial motion,  is expressed as:

\begin{equation}
  \dot{M}_{B} \equiv \pi e^{3/2} \rho_\infty \frac{G^2 M_{BH}^2}{c_{\infty}^3},
\end{equation}

where an adiabatic index $\gamma=1$ was assumed. The Bondi rate
can be much larger than the Eddington rate given sufficiently high density and temperature, but
also significantly smaller, as in the case of mini-halos discussed in Chapter 10.
As the temperature in galactic nuclei should not be much different from the mean temperature of the
galactic ISM, namely in the range $10^3-10^4$ K (molecular gas would be colder but it should be
dissociated efficiently by the warm ionizing flux resulting from accretion onto the Bh seed), the
ambient gas density $n_\infty$ becomes  the most important variable. For example, for light
BH seeds formed by Pop. III remnants, with masses around $100 M_{\odot}$, the corresponding density
must be $> 10^7$ atoms/cm$^2$ for the above condition to be satisfied.
Qualitatively, the condition of a dense, highly optically thick flow yielding very long 
photon diffusion timescales reminds of the conditions in the SLIM disk model, except that here there is 
no role of advection. Furthermore, the seemingly high gas densities required to initiate sustained
Super-Eddington accretion are commonly found in the center of protogalaxies or circumnuclear regions
of high-z gas rich galaxies in hydrodynamical simulations \citep[see e.g. next two sections and][]{Lupi16}.

\citet{Takeo18} have improved further such model by studying the effect of the anisotropy of the 
radiation field emitted around the black hole, to mimic the effect of an unresolved accretion disk,
for which one expects radiation to be primarily emitted around the rotation axis of the disk \citep{Ohsuga2005,Jiang14,Sadowski16}.
The starting point is still a spherically symmetric gas cloud, and a large-scale gas infall rate is
given as a boundary condition. They used  a 1D multi-frequency radiative transfer code coupled with 
a 2D hydrodynamical simulation and a non-equilibrium primordial chemistry network. Self-gravity of the
gas cloud, assumed to be initially of uniform density, 
was neglected, and a finite difference approach to solve the hydro equations was employed. The cloud
is initially static.
They consider a range if BH seed masses ($10^3 - 5 \times 10^5 M_{\odot}$) along with 
different geometries for the radiation field. The radiation field geometry was chosen as an {\it ansatz} in the 
initial conditions by projecting the radiative flux vector over a fraction of the solid angle. The
underlying assumption is that the accretion flow will be organized in a disk or torus, or, more generically,
would have an anisotropic pattern, thus leaving funnels of low density gas along which radiative losses 
would preferentially occur. Furthermore,  only  the radial
component of the radiation force was considered.
Because of anisotropic emission, matter was found to accrete preferentially
in the plane perpendicular to the emission direction. The main difference with the calculations for isotropic
radiation fields is that, for the same envelope density, lower mass BH seeds, down to $10^3 M_{\odot}$, can enter a sustained
super-critical accretion (see left panel of Figure 5). 
This is interesting since this is in the mass range of Pop. III remnants. 
Another
novel result found by such calculations is that, as the mass of the BH seed increases, a transition is observed
in the qualitative evolution of the gaseous envelope. Indeed above a BH mass of $5 \times 10^5 M_{\odot}$ 
the ionized gas around the
hole collapse due to super-sonic non-radial motions of neutral gas and radiative recombination. The neutral gas thus absorbs
the momentum carried by photons and
is accelerated outward in funnels along the poles
triggering warm outflows with $T \sim 8000$ K. The latter finding could interesting observational implications that may
tell the model apart from other that have been proposed for Super-Eddington accretion.

\begin{figure*}[h]
\includegraphics[width=5 cm]{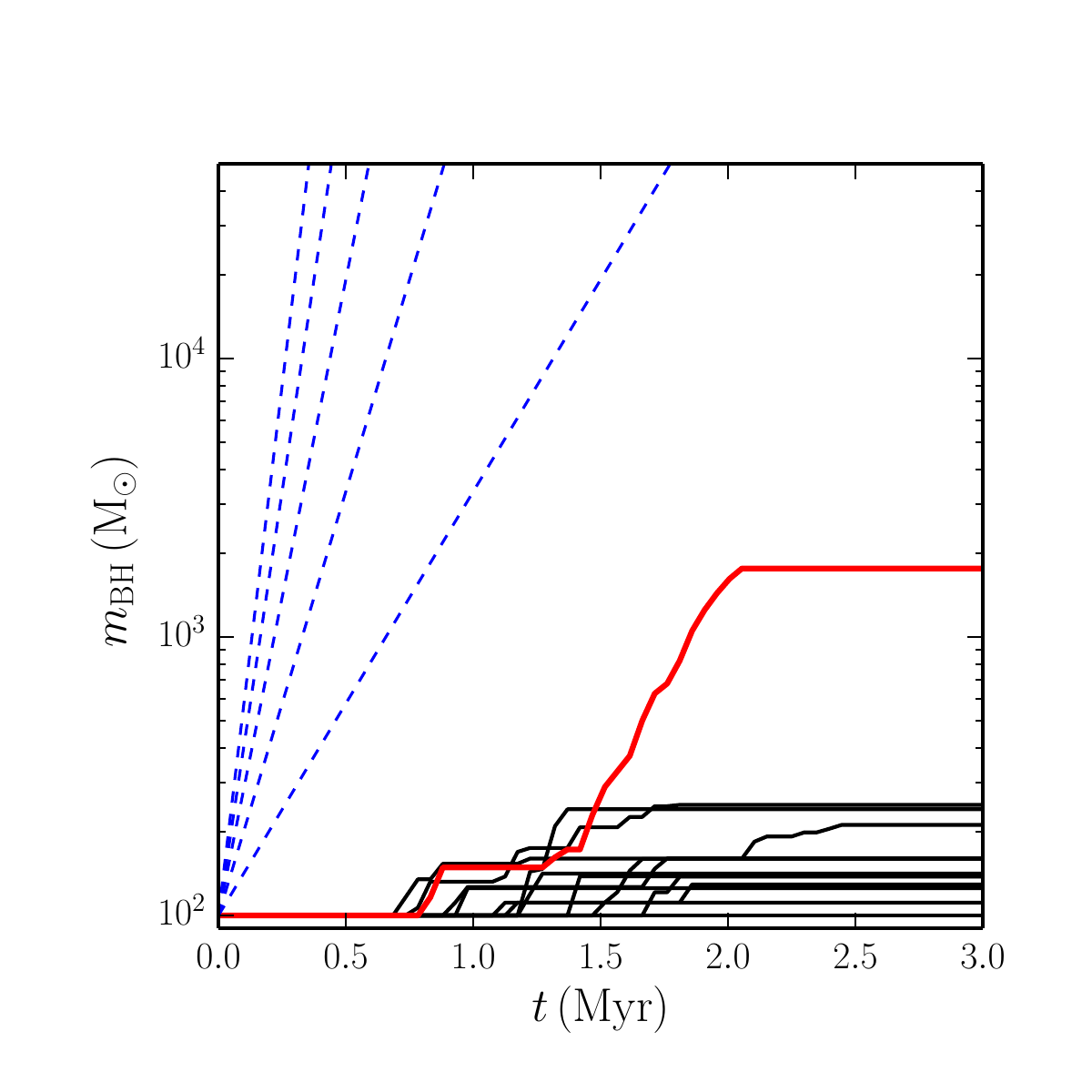} \hspace{0.5 cm}
\includegraphics[width=5 cm]{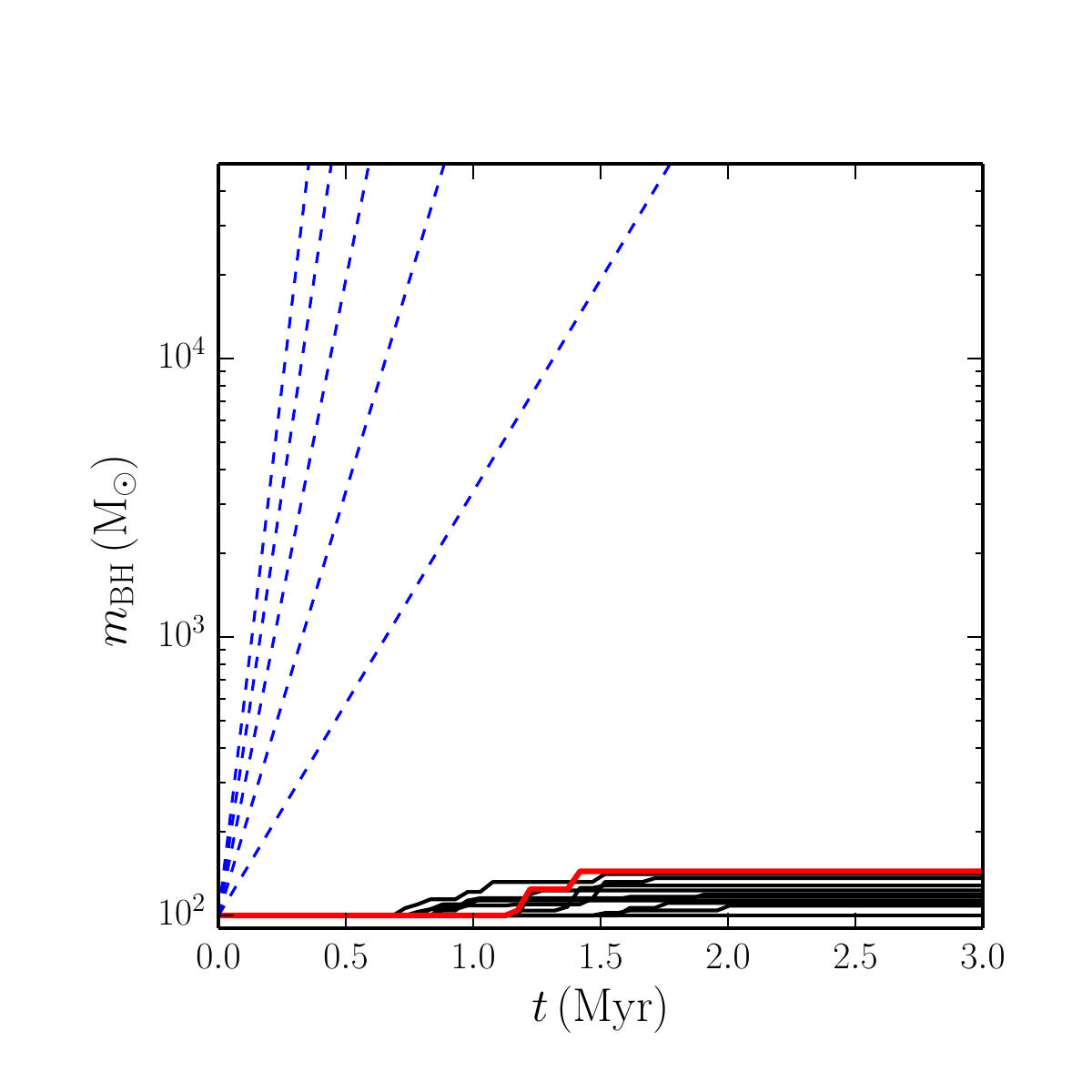}
\caption{\label{label} Mass growth of light black hole seeds in a simulation using the SLIM disk model for accretion  
(left panel) and in one
using the standard $0.1$  radiative efficiency for a viscous disk (right panel), from \citet{Lupi16}. 
The red lines correspond to the most massive BHs at the end of the runs,  the black lines correspond to all other BHs
in the CND,  while the blue dashed lines depict reference accretion histories 
at fixed Eddington ratios of 500, 400, 300, 200, and 100,  respectively. 
The marked difference in growth rate and absolute growth between standard and SLIM disk-like radiative
efficiency is evident. Figure adopted from \citet{Lupi16}, reproduced by permission of Oxford University Press / on behalf of the RAS.}
\end{figure*}

\section{Large scale flow conditions in primordial galaxies: Super-Eddington gas supply rates?}

In order to sustain Super-Eddington accretion an necessary condition is that
the accretion disk/envelope itself is fed at its edge from large scales with gas impinging at  very high rates, as assumed
indeed in the \citet{Jiang17} and \citet{Sadowski16} simulations.
We can ask what is the current evidence of high inflow rates at scales just
above the accretion disk. We can also ask if such flow has a high angular momentum or not,
and thus whether or not accretion will occur indeed through a disk-like interface.

First of all let us set the spatial scale of interest. While that specifically will
depend on the mass of the MBH under consideration, we can proceed
conservatively and decide that any result concerning the flow
properties below pc scales, possibly below $0.1$ pc, would
be relevant as it would provide the natural boundary condition
for smaller scale models/simulations starting at the edge of the
accretion disk. Indeed the accretion disk around an MBH of a few thousand solar mass
or larger, hence ranging from the mass of a Pop. III remnant to the mass
of a full-fledged MBH, will have a physical radius in the range  $10^{-3}$ and $10^{-2}$
pc roughly. 
The most interesting
case is that of an MBH weighing the few hundred to a few thousand solar mass,
which would be the expected mass for the remnant of a Pop. III star. The
latter scenario has been considered in recent works modeling accretion
from a spherically symmetric envelope as in Inayoshi et al. (2017).

In the remainder we will summarize the gas inflow rates in the range kiloparsecs
to $0.1$ pc found in  simulations of protogalaxies and high-z galaxies.
First we cover the results concerning metal-free protogalaxies, namely galactic-scale
objects forming in mini-halos with a virial mass $< 10^{10} M_{\odot}$, most typically in the range
$10^8-10^9 M_{\odot}$, at $z \sim 15-20$. These are the typical target of the simulations
designed to study the possibility of massive BH seed formation by direct gas collapse
in metal-free halos, which were thoroughly described in Chapter 5 of this book.
In this case inflow rates are in the range
$0.1-1 M_{\odot}$/yr (Latif et al. 2016) at scales of $0.1$ pc. If we assume that
a light BH seed would be present in such a protogalaxy, such inflow would be well above the Eddington limit
for BHs of a hundred to several thousand solar masses.
Indeed the accretion rate at the Eddington limit, $\dot M_{Edd}$ can be conveniently
expressed as:

\begin{equation}
\dot M_{Edd} = (4 \pi G / \epsilon \kappa c) M_{BH} = 2.2 \times 10^{-8} (M_{BH} / {M_{\odot}) M_{\odot} yr^{1}}
\end{equation}

from which it is clear that the Eddington limit for a BH with mass of $1000 M_{\odot}$ corresponds to $2.2 \times 10^{-5} M_{\odot}$/yr.
In the equation above we have assumed an opacity $\kappa = 0.4$ cm$^2$/g and $\epsilon = 0.1$, as in standard Eddington-limited accretion
flows. However, Super-Eddington accretion models discussed in the last section predict rates that are in the range
a few tens to several thousand $M_{Edd}$ for both super-critical flows in accretion disk and in extended envelopes
(see section 2). For a BH of $1000 M_{\odot}$ these numbers would imply an inflow rate of $10^{-3}-0.1 M_{\odot}$yr${-1}$
to feed the nuclear region at the corresponding rate, which are still well within the range of the pc-scale inflow rates in the
simulations of metal-free protogalaxies.

\begin{figure*}[h]
\includegraphics[width= 5 cm]{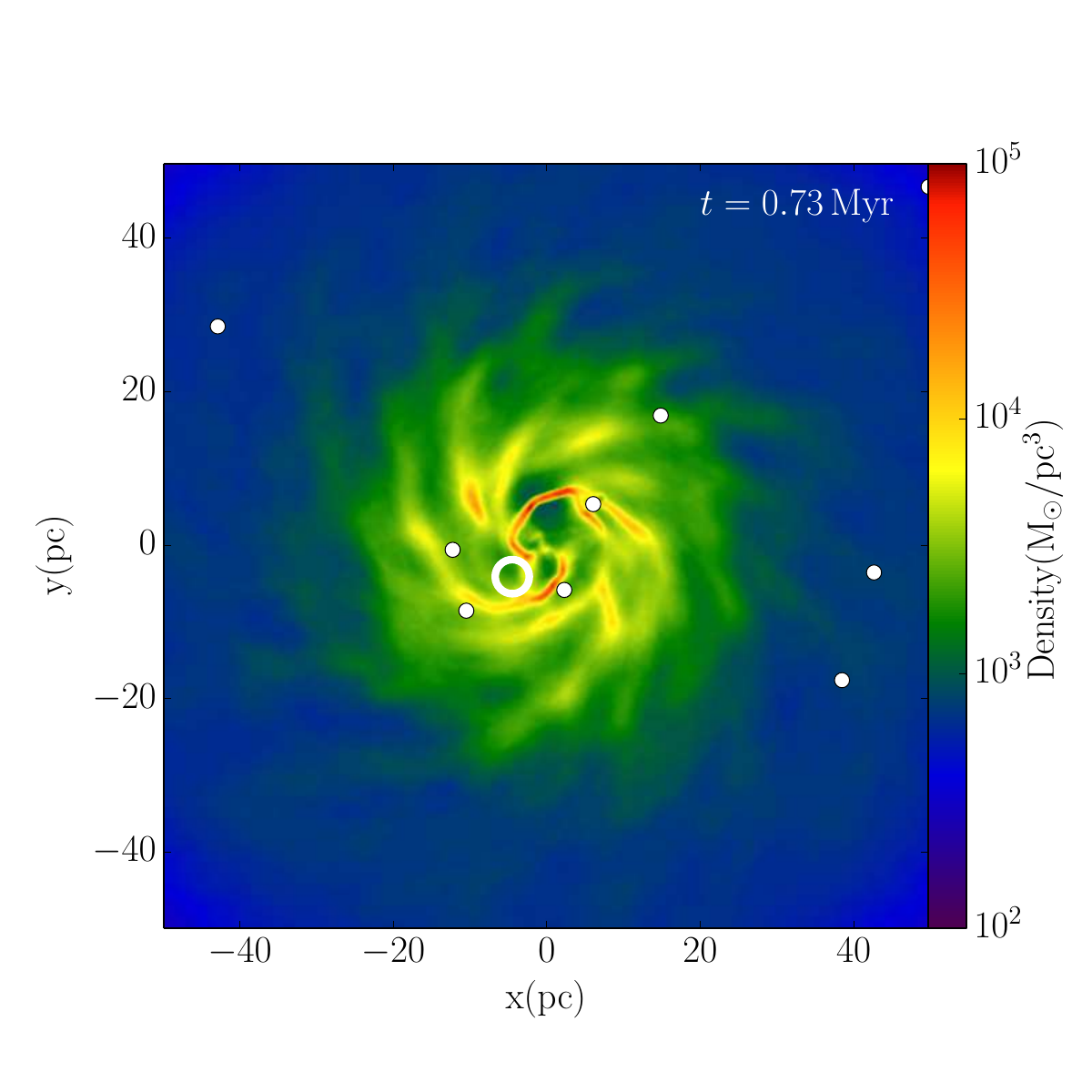}
\includegraphics[width=5 cm]{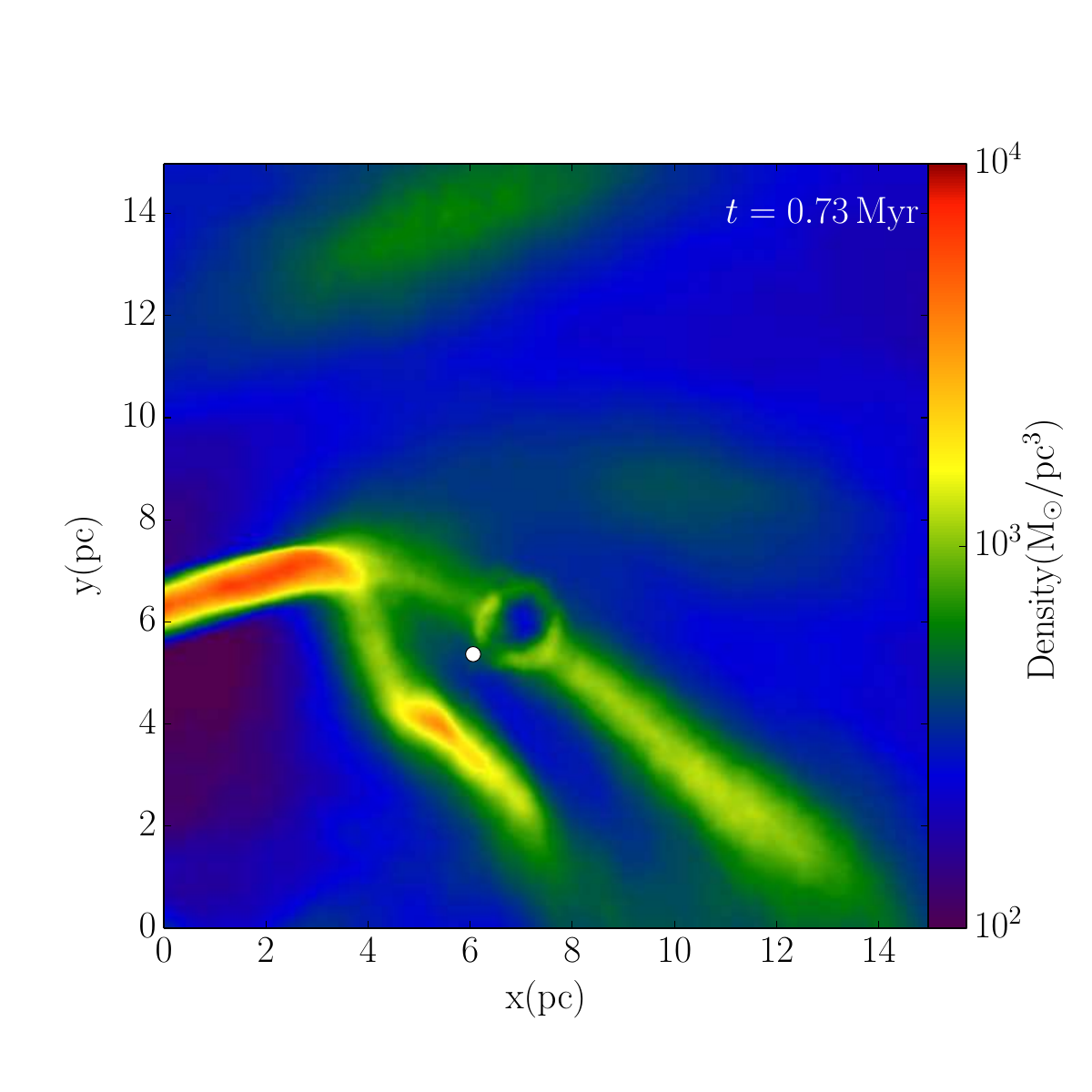} \\
\includegraphics[width=5 cm]{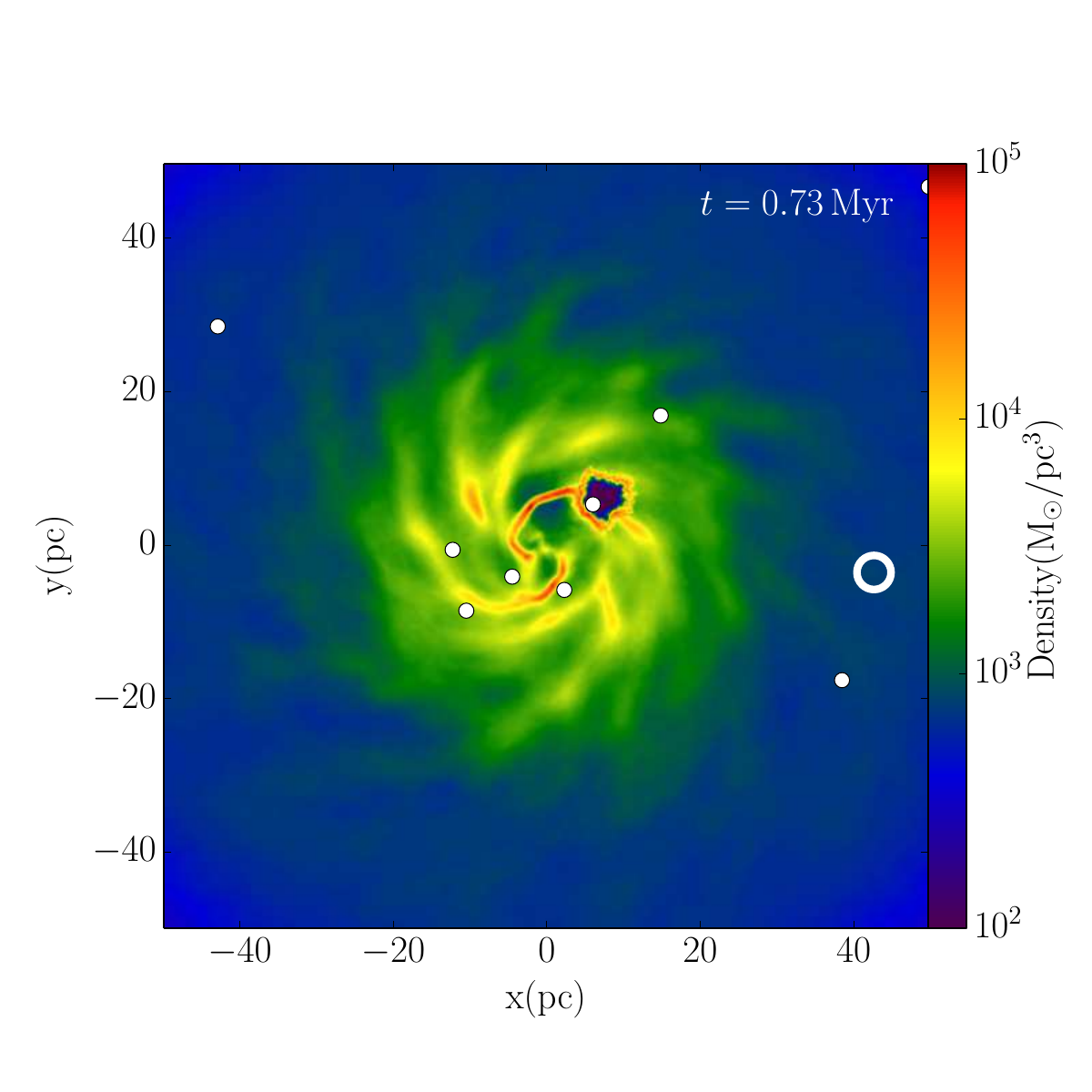}
\includegraphics[width=5 cm]{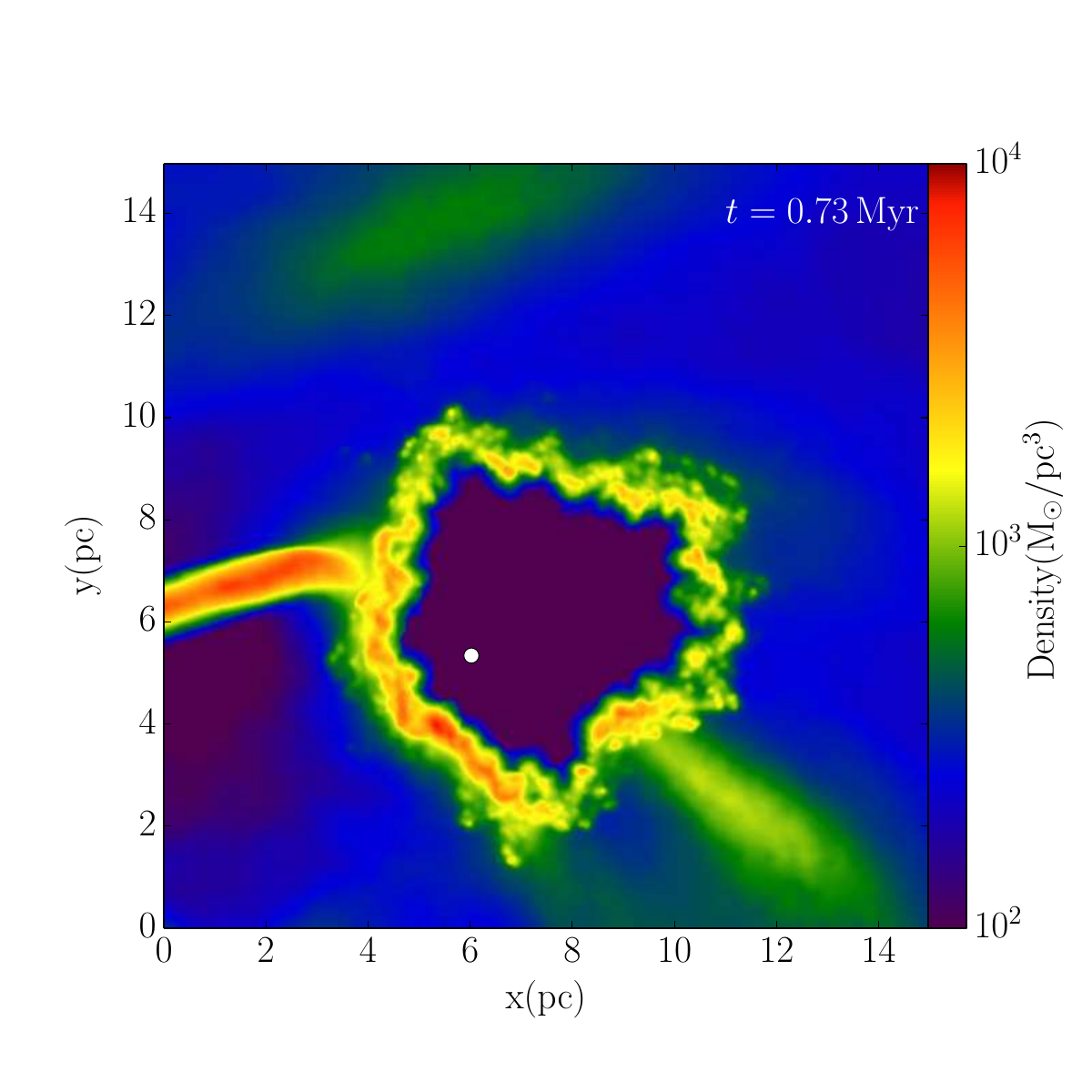}
\caption{\label{label}
 Density maps of the circumnuclear disk (CND) at large (left) and small scales  (right) in simulations
carried out with a SLIM disk sub-grid model (top) and with a standard $0.1$ radiative efficiency as expected
in conventional viscous disk accretion (bottom), from \citet{Lupi16}.  These are the same runs analyzed
in Figure 7. The maps highlight the
different effect of BH (radiative) feedback in the two cases. Indeed in the SLIM disk case, due to 
much lower radiative efficiency, the bubble created by feedback around the BH is much smaller
and weaker (top-right panel), promoting faster sustained growth relative to the standard accretion case (bottom-right
panel). Figure adopted from \citet{Lupi16}, reproduced by permission of Oxford University Press / on behalf of the RAS.}
\end{figure*}

The mass inflow rate of gas with negligible pressure support pulled in by the gravitational action of the
halo, and neglecting the effect of angular momentum, is $\dot M \sim \alpha V_{inf}^3/G$ , where $V_{inf}$ is the
infall velocity and $\alpha$ is the coefficient of viscous
drag relenting the inflow in an eventual disk-like envelope \citep{Mayer18}.
To first order inflow rates should thus scale as  $\alpha V_{vir}$, where we have taken the characteristic infall 
velocity to be of order the virial velocity  scale of the halo. 
Note $\alpha \sim 0.01-0.1$ if self-gravitating instabilities
are the main transport mechanism  and they are treated as a local 
effective viscosity acting on the flow \citep{Lodato04}. By assuming $\alpha = 0.01-0.1$ one recovers
inflow rates in the range $0.1-1 M_{\odot}$/yr, comparable to those in the protogalaxy simulations \citep{Latif2013c,Latif2015}, which have halos masses $\sim 10^9 M_{\odot}$.
Based on the scaling $\dot M \propto M_{vir} \sim {V_{vir}}^3$ one expects much higher rates in more massive halos,
100-1000 times larger indeed in $10^{10} - 10^{11} M_{\odot}$ halos, hence in the range 100-1000 $M_{\odot}$/yr.
Equation (4) indicates that an inflow rate
of 1000 $M_{\odot}$/yr would be super-critical even for a $10^8 M_{\odot}$ BH, suggesting that in these
larger galaxies super-critical accretion is possible throughout all stages of black hole growth, across 
many decades in mass.
Halos with virial masses in the range $10^{11}-10^{12} M_{\odot}$ appear below $z=12$, hence in an epoch
in which gas in such biased halos will be already metal-enriched, rendering radiative cooling is more efficient, which also can boost
gas inflow further  \citep{Mayer2015}. 
Both observations, which are just beginning to probe this phase \citep{Grazian15,Vito18} and simulations \citep{Feng14, Feng16}, agree on the point of fast metal enrichment in such biased
high density peaks at high-z.

The MassiveBlack  \citep{DiMatteo2012}, MassiveBlackII  \citep{Khandai15}, and BlueTides \citep{Feng16}
simulations have studied gas infall in these massive
high-z galaxies with the goal of assessing the growth history of putative BH seeds at their centers.
All these simulations modeled large cosmological volumes in order to be able to identify highly biased density
fluctuations that should correspond to the hosts of high-z QSOs \citep{Li06}. 
Given the large volumes adopted they limited their analysis to scales of order a kiloparsec, corresponding to their 
resolution limit. Due to the same resolution limit they assumed initial BH seeds with already large masses, 
$> 10^5 M_{\odot}$.
Therefore, they mainly assessed the rates of gas infall from the cosmic
web onto the nascent galaxies in very massive halos. \citet{DiMatteo2012} were the first to show
that the feeding occurs mainly in the form of cold gas flows ($T \sim 10^4$ K) accreting directly from 
the cosmic web, although a circumnuclear disk, perhaps unstable and clumpy, could be present at smaller
unresolved scales \citep{Gabor_Bournaud_2013}.
Infall rates of order $100 M_{\odot}$/yr sustained on Gyr timescales were observed in these simulations, even in the
presence of AGN feedback (starting from an initial BH seed with mass $\sim 10^5 M_{\odot}$  implanted in each halo). 
While in the simulations the sub-resolution accretion onto BH seeds was capped at the Eddington limit (except in the BlueTides
suite, in which it was allowed to reach 3 times the Eddington value, see \citet{Feng16} these prominent
inflows would be highly Super-Eddington if they could continue unimpeded all the way to edge of the accretion disk/
envelope.
Whether or not they can really continue unimpeded to smaller scales will depend
on the gas dynamics inside the galaxy, hence on the nature of the galactic potential and of the density distribution.

Only very recently it has become possible to tackle directly this question using the zoom-in technique by increasing
many-fold the mass and spatial resolution in a selected object within a large cosmological volume. 
The MassiveBlackHR simulation (Capelo, Mayer, Di Matteo \& Feng, in prep.) achieves this in {\it Halo 3} of the precursor MassiveBlack
cosmological volume. The selected halo has clustering properties
consistent with those of bright $z \sim 6$ QSOs \citep{DiMatteo2012}.
The simulation, which employed the unprecedented resolution of 500 million fluid elements (SPH 
particles in this scale) reaches a spatial resolution of a few tens of parsecs in the gas
phase. Preliminary analysis shows that a prominent gas disk a few kpc in size forms at the center of the main
dark matter halo at $z \sim 8$, following a merger between two massive proto-galaxies. 
The disk, with regular grand-design spiral structure and a central bar-like distortion, is 
reminiscent of low-z disks in galaxies, has a mass comparable to the disk of our own Milky way  but it is a factor of 10 
more compact size, has a much higher gas content relative to the stellar content (50\%) and thus a much higher density. 
This simulation shows
that not only high gas inflow rates ($> 100 M_{\odot}$/yr)
persist down to sub-kpc scales, but they occasionally increase further,peaking at
$1300 M_{\odot}$/yr, mostly in coincidence with mergers and flybies.
Vigorous bars and spiral density waves do aid angular momentum transport via non-axisymmetric
torques at sub-kpc scales, confirming the findings for much smaller simulated protogalaxies at $z > 10$
(Chapter 10). If such inflows continue to pc scales at similar rates they would, of course, be highly Super-Eddington accretion.


We have seen that massive halos with $M > 10^{10} M_{\odot}$ are in principle capable of sustaining
super-critical accretion over a range of masses, from light Pop. III seeds to black holes with masses
$\sim 10^8 M_{\odot}$.  There is, though,  another important consideration that goes in the direction of favouring halos with masses 
above $10^{10} M_{\odot}$ as natural environments
for super-critical accretion; this is  the effect of stellar and SN feedback as
a function of halo mass. At moderate redshifts, $z  < 5$,  
it is now fairly established that dwarf galaxies with a virial mass $M_{vir} 
< 10^{10} M_{\odot}$  cannot preserve
large amounts of dense gas in their nuclei because of the strong effect of winds/outflows driven
by SNe  \citep{Governato10,Shen14,Tollet16,Habouzit17},  see also chapter 10. While no specific studies have been carried out to explore the detailed conditions of
nuclear gas flows in realistic high-z dwarf galaxy hosts, one expects a qualitatively similar behaviour.
Moreover, feedback-driven outflows should be widespread in low mass galaxies and become even stronger
at high redshift reflecting the notion
that specific star formation rates increase at a fixed galaxy stellar (as indeed verified in the
simulations of  \citet{Fiacconi_et_al_2017a}).
As for mini-halos at even higher-z, episodic
accretion of the central BH would contribute further to heating and expelling the gas (see Chapter 10).  
Therefore the consistently large 
gas inflows necessary for sustained Super-Eddington accretion should be more prevalent
in such galaxies hosted in  massive halos since these  are
more resilient to the effect of feedback-driven outflows. We also note that, typically, in simulations
of high-z metal-free protogalaxies (see Chapter 5) {\it normal}
star formation beyond the stage of Pop. III stars, and thus its resulting feedback
is not taken into account, hence the the effec of feedback is underestimated.
It follows that halos with $M_{vir} > 10^{10} M_{\odot}$ are
the natural candidates to support sustained super-critical accretion onto their central BHs. 

Episodic Super-Eddington 
accretion might occur for a wider
range of halo masses, especially in the presence of strong tidal forces promoting temporarily gas inflows 
during mergers (major and minor), or in strong tidal interaction events, all particularly frequent
at high-z.
Episodic intense accretion might still be capable of growing some BHs
to large sizes even starting from lights seeds, especially in highly biased regions such
as those that will later host bright QSOs since these are by construction growing
their mass faster \citep{Pezzulli16,Pezzulli17}. \citet{Pezzulli17} found only a 50\% decrease in the overall population of MBHs that
grow efficiently via super-critical accretion at $z > 7-8$ in their semi-analytical model when the effect of
stellar and SNe feedback are taken into account. However it is not known how much bigger such effects 
would be in halos with $M_{vir} < 10^{10} M_{\odot}$ since the latter work focuses on halos that reach
a mass as large as $10^{13} M_{\odot}$ at $z=\sim$.
In any case, in their semi-analytical
model mergers play a major role in revitalizing inflows even in low mass halos, an aspect that should
be tested with detailed high-resolution cosmological hydrodynamical simulations.

\section{Models of Super-Eddington growth of massive BH seeds in high redshift
galaxies}

In principle the growth of BH seeds in the early Universe should be studied
using multi-scale  simulations that capture the accretion disk flow as well
as the larger scale feeding of gas from the surrounding interstellar medium
to the disk. However this is not computationally feasible at the moment.
as the range of densities and timescales is prohibitive. The difficulty is
exacerbated further by the need
of treating radiative processes at all scales, and eventually MHD and relativistic
effects at the scale of the accretion disk. This is a daunting task with
current hydro codes even on large parallel supercomputers as the multiple timescales
introduce load balancing bottlenecks that prevent efficient scaling on large
node counts. Already at the scale of the accretion disk itself the inclusion
of advanced radiative transfer schemes in a global 3D simulations, such those of \citet{Jiang17} 
described in section 2, results in a computational challenge.
Therefore connecting such simulations
to the protogalaxy simulations starting from cosmological initial conditions, the topic
of the previous section, requires
some modeling interface to simplify the computation. This interface can be realized
in the form of a sub-grid model of disk accretion applied to the BH seeds that
evolve in simulations of the protogalactic environment. Similar sub-grid approaches are
routinely adopted to study the growth of MBHs at lower
redshift in cosmological simulations, which describe mass growth as well
as radiative and momentum feedback from the accreting MBH onto the
surrounding ISM (the so-called AGN feedback, see Chapter 10).

The main difference is that in conventional cosmological/galaxy-scale simulations accretion
rates are capped at the Eddington limit \citep{DiMatteo_et_al_2005,Hopkins:2011p2034},
whereas in this Chapter we want to discuss how the gas flow behaves when one drops such constraint.
The SLIM disk model lends itself to be cast in the form of a sub-grid model given
its relative simplicity, in that its main features can be encapsulated in the
resulting radiative efficiency (see section 2). This 
indeed has been done in \citet{Lupi16}, who have studied the growth of light BH
seeds in the nuclear regions of protogalactic disks using hydro simulations by allowing
accretion to be come super-critical, as well as in \citet{Pezzulli16,Pezzulli17}, who have
modeled super-critical accretion in a semi-analytical model of galaxy and black hole
formation.
In a preceding work by \citet{Madau2014} the
same SLIM disk model has been used in a semi-analytical code to show 
that even very light seed BHs originating from Pop. III stars can grow to
billion solar masses by $z \sim 6-7$. In such work the seed BH was considered to be
isolated and with an arbitrary large gas reservoir to feed onto.
This was shown
to be the case for both steady and intermittent super-critical accretion. The
accretion rate was only a few times above Eddington (Figure 6), which is a very conservative
assumption in light of the results of the small-scale simulations described in section 2.
Madau et al (2014) proposed the following
spin-dependent fitting equation to the numerical results of \citet{Sadowski9}:

\begin{equation}
L/L_E=A(a) \left[ \frac{0.985}{\dot m_E/\dot m+B(a)}\,+\,\frac{0.015}{\dot m_E/\dot m+C(a)}\right],
\end{equation}

where the functions $A, B,$ and $C$ scale with the spin of the black hole as

\begin{eqnarray}
A(a) & = & (0.9663-0.9292a)^{-0.5639},\\
B(a) & = & (4.627-4.445a)^{-0.5524},\\
C(a) & = & (827.3-718.1a)^{-0.7060}.
\end{eqnarray}

Because photon-trapping occurs in Super-Eddington accretion the emitted luminosity is not
linearly proportional to the accretion rate anymore, \citet{Madau2014} also propose a
convenient way to recast the accretion timescale equation (see above equation (1)):

\begin{equation}
t_{\rm acc}={t_E\over 16(1-\epsilon)}\left({\dot m_E\over \dot m}\right) (8.4\times 10^6 {\rm yr})\,\left({3\dot m_E\over 
\dot m}\right),
\label{eq:tsacc}
\end{equation}

where the last inequality holds for moderate Super-Eddington rates independent of the value of the black hole spin
 \citep{Madau2014}. 
While low radiative
efficiency of the SLIM disk model is accretion-rate dependent, the equation shown
above shows that it is very mildly dependent
on the spin, at stark variance with what occurs for the standard viscous disk model for
which spin dependence is important. The next key question is whether or not the conditions
for Super-Eddington accretion, namel sustained high gas accretion rates, are physically
possible in protogalaxies. \citet{Lupi16} indeed addressed this aspect using idealized
simulations that only model the circumnuclear region of dense gaseous protogalaxies (the circumnuclear
disk, hereafter CND). This approach allows to reach very high
resolution and thus study the nature of the gas flow at scales even as small as 0.02 pc. Note,
though, that this
is still an order of magnitude larger than the sphere of influence of a stellar black hole coming from the collapse 
of even the largest Pop. III stars (of order a few thousand solar masses).
Controlled experiments with isolated systems allow to survey the parameter space extensively
at high resolution, which would require prohibitive computational costs with cosmological simulations.
\citet{Lupi16} performed their simulations using two different numerical hydrodynamics techniques,
the Lagragian finite mass meshless method (MFM) in the GIZMO code  \citep{Hopkins13} and
the adaptive mesh refinement code RAMSES  \citep{Teyssier2002}.
Star formation and feedback from Supernovae were included, adopting the so-called blast wave
model by  \citet{Stinson06} in both codes, in which cooling of the gas heated by SN ejecta is temporarily
suspended to mimic unresolved turbulence and momentum deposition. Black holes were treated
a massive collisionless particles that accrete at a rate determined by their radiative efficiency.
Initially black  holes with a range of masses were distributed at random locations inside the CND. 

Generally speaking there are two ways by which a low radiative efficiency can affect
BH growth. First of all, for a given accretion rate BHs grow faster with lower radiative efficiency because
less mass is converted into radiant energy (by a factor proportional to $1 - \epsilon$). The second way is that
BH feedback is reduced with a lower $\epsilon$, once again because the radiant energy emerging from the
accretion process is lowered, which implies a larger accretion rate can be sustained. The mechanical
component of feedback manifests itself as momentum-driven winds and/or outflows. Its strength
does not have a natural scaling behaviour with radiative efficiency, but still has to decrease because of the first effect. 
Feedback onto the surrounding 
ISM by the accreting  black hole was included in \citet{Lupi16} by adopting the thermal coupling model in
\citet{BoothS09}, hence neglecting momentum feedback. \citet{Lupi16} then compared the results obtained with standard viscous disk radiative efficiency ($\epsilon = 0.1$)
against those with a reduced efficiency based on the SLIM disk model, and found 
striking differences concerning the growth rate
of the BH seeds. Furthermore, black holes were also allowed to merge, which introduced another
channel for growth. The resulting scenario is thus more complex compared to the toy model presented in  
\citet{Madau2014},
in which only growth by gas accretion of a single, isolated black hole seed was considered. 
Furthermore, as the CNDs were dense and massive  
self-gravity played an important role, causing fragmentation into clumps which changed the nature of the
gas flow (see Figure 8).
These very dense clouds were weakly affected by feedback owing to their high density (much higher than
present-day Giant Molecular Clouds (GMCs)), and even more so in the case of the SLIM disk accretion regime.
This clumpy flow favoured a sustained accretion rate, quite irrespective of hydro method and resolution.

\begin{figure}[h]
\includegraphics[width=34pc]{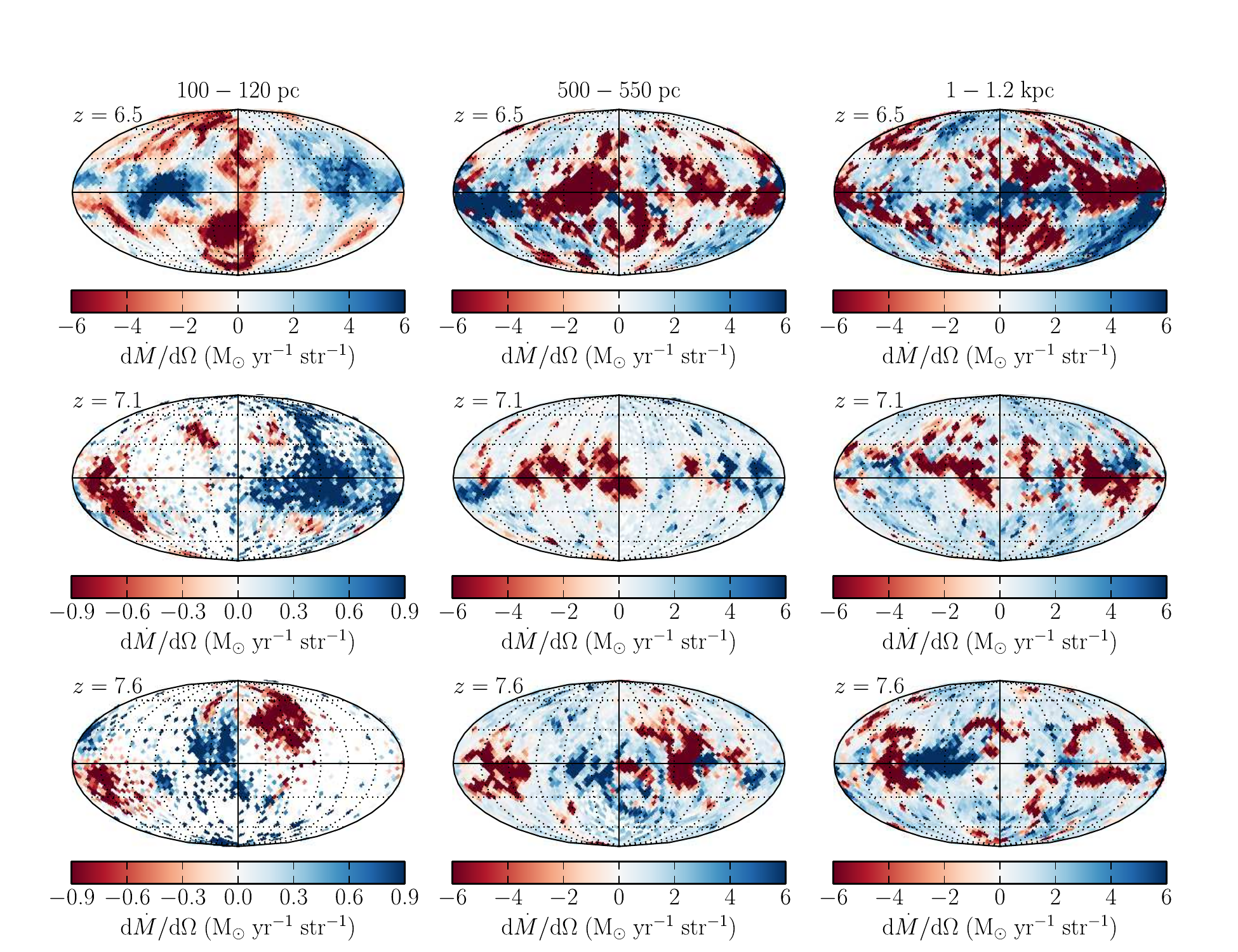}\hspace{2pc}
\caption{\label{label}Mollweide projections of gas density through the center showing the inflow and
outflow to/from the nucleus at different scales (left  to right) and different redshift 
(from bottom to  top redshift decreases) for the "PONOS"
cosmological galaxy formation simulation  of \citet{FiacconiMayer17}. The gas inflow pattern is anisotropic at all scales, down to the
nuclear region. The hydrodynamical resolution of the simulation  reached a few pc but the gravitational softening was
set to 40 pc. Figure adopted from \citet{FiacconiMayer17}, reproduced by permission of Oxford University Press / on behalf of the RAS.}
\end{figure}

Despite the intrinsic complexity, \citet{Lupi16} found a recurrent outcome in their simulations, namely
that one black hole, which had experienced faster growth since the beginning, was overtaking all the
others by capturing competitively most of the available gas in the surroundings as well as  merging with the other black holes. This recalls 
the behaviour in competitive accretion in star formation \citep{Bonnell01}. 
Growth was typically
terminated when all the gas was consumed, predominantly into star formation rather than accretion 
(the mean star formation rate in a sphere of 1 pc was $0.1 M_{\odot}$/yr as opposed to $10^{-3} M_{\odot}$/yr
for the black hole accretion rate). We remark in the simulations there was no gas infall onto the CND from
exterior regions
that could prolong sustained growth of the black hole seeds.
By the end of the simulations, which extended for 3 Myr, when the gas in the CND was depleted,
the mass of the major seed BH reached values $10^4-10^5 M_{\odot}$ in the SLIM disk case, while it barely
reached $10^3 M_{\odot}$ when using standard disk accretion efficiency (see e.g. Figure 7). Remarkably, 
the accretion rates in the SLIM disk case were in the range 100-200 times the Eddington mass accretion rate,
which is consistent with the results of the accretion-scale simulations described in section 2.
The most massive BH will inevitably migrate to the center via dynamical friction on a few orbital times,
explaining why MBHs are found at the center of galaxies.

Finally, it is important to remark that these CND simulations, having a resolution of $0.1$ pc, 
do not resolve the scale of the accretion disk, hence it is not yet possible to connect the gas flow
properties at the numerical accretion radius to the gas flow in the MHD disk simulations of
Super-Eddington accretion discussed in section 2. Nevertheless, one can measure the angular momentum
of the gas that accretes on the BH sink particles. It was found that this is very low, as most of the
gas accretes in the dense clumpy phase, namely in a regime in which the gas flow in the CND by nature highly
asymmetric and chaotic. As a result in the regime of these simulations
accretion, while not spherical and isotropic, ends up having a rate comparable to the Bondi rate
for the temperature and density of the local medium around the BHs. Whether this is realistic or not has to be ascertained with cosmological
simulations, which provide realistic boundary conditions for the flow in the CND,as well as with
improved radiation physics, such as implementing the self-shielding of gas and the local photoionization,
which could both suppress cooling, and thus fragmentation. 

\citet{FiacconiMayer17} made a first step in addressing the problem  of the realism of the initial
conditions by studying the nature and geometry of the gas flow in the nuclear region of a $z \sim 5-6$
massive galaxy in the state-of-the-art cosmological hydrodynamical simulation dubbed "PONOS". They found
that the mass of the region within a few 100 pc from the center was varying in the range $10^7-10^8 M_{\odot}$,
relatively comparable with the mass assumed in the CND models of \citet{Lupi16}. However, gas inflow
rates were observed to be highly variable and anisotropic down to the sub-kpc scale of the nucleus (Figure 9). 
This is a generic expectation given the highly nonlinear and chaotic gas accretion pattern from the cosmic
web to the disks, and is confirmed also in the simulations of more massive $z \sim 7-10$ protogalaxies in the MassiveBlackHR
run (Capelo et al, in prep.).
This, perhaps combined with the resolution limit set by the gravitational softening of 40 pc,  did not produce a long-lived CND-like
structure as that assumed in \citet{Lupi16}.
Another potential difference was that the effect of stellar and SN feedback
was found to be much stronger than in the CND-scale simulations of \citet{Lupi16},
suppressing fragmentation and leading to a turbulent flow from tens of pc to kpc scales.
The highly anisotropic and turbulent nature of the flow down to the nuclear region is qualitatively consistent with the
fact that a coherent structure such as CND is not assembled  (Figure 9).
While the long term evolution
of the angular momentum in the turbulent nuclear region was not studied, dissipation through shocks between 
colliding flows of turbulent eddies is expected to be capable of extracting angular momentum and sustain
accretion to smaller (unresolved) scales. The mean inflow rates on scales of a hundred pc are indeed still high enough to be super-critical
for a BH seed sitting in the center. The proposed chaotic accretion scenario \citep{King06} might apply in this case.

\section{Summary}

We have reviewed the status of knowledge relevant to Super-Eddington accretion of BH seeds at high redshift.
The latest developments of small scale  simulations with detailed physics modeling accretion disks and envelopes
in super-critical regimes were presented and discussed in the context of massive BH seeds. We the covered gas inflows rates
in protogalaxies at $z > 15$ as well as in more massive galaxies in high sigma peaks at $z \sim 10$ and below.
This is relevant to determine how efficiently the accretion disk or envelope is fed with fresh fuel from the
surrounding ISM, and in particular if the inflow rate is compatible with the accretion rates found
in small scale simulations of  super-critical phases. We then concluded by describing results of simulations that
attempt to bridge the scales by treating super-critical accretion with a sub-grid model. Currently only the
SLIM disk solution has been considered in such sub-grid models. The main findings that we discussed are:

\begin{itemize}

{\item Super-Eddington accretion in disks and envelopes is feasible and, if it happens, it reaches high amplitudes, from ten times 
to a few  thousand times above the Eddington limit. }

{\item The radiative efficiency characterizing optically-thick, super-critical flows is low, but whether it is 10 or 1000
times lower than the standard radiative efficiency in the conventional thin "$\alpha$" disk model is
still unclear, and seems to depend on the method adopted by simulations to solve for radiation transport.}

{\item The ratio between radiative and kinetic luminosity emitted in a Super-Eddington accretion phase is also
uncertain. In some cases it makes up for most of the emitted energy, which has important implications on how
to model feedback from the BH seed or MBH onto the surrounding interstellar medium}

{\item Pc-scale gas inflow rates in  protogalaxies at $z \sim 15-20$ or the first massive galaxies at $z < 8-10$
are always above the Eddington limit for light BH seeds ($M_{BH} \sim 100-1000 M_{\odot}$),  but can be such for a 
wide range of black hole masses, up to $10^8 M_{\odot}$, only if we consider galaxies in high sigma peaks
with halo masses $10^{11} - 10^{12}  M_{\odot}$. Only in this latter case prolonged Super-Eddington accretion
appears feasible.
Furthermore, galaxies in low mass halos might not have a long-lasting nuclear gas reservoir due to the
strong effect of stellar and SN feedback, which might stifle BH growth via Super-Eddington accretion.}

{\item The results of the SLIM disk accretion model can be easily parametrized as a (very low) modified radiative efficiency,
with a weak dependence on the black hole spin. With the latter recipe both semi-analytical models and local, non-cosmological 
simulations of circumnuclear disks (CNDs) have obtained encouraging results, showing that super-critical accretion can
occur steadily as well as episodically depending on the environmental conditions
but always leads to rapid growth of light BH seeds, matching naturally the constraints
imposed by the rapid emergence of high-z QSOs. However it is still unclear if the dense CND assumed in
such simulations occurs in a realistic cosmological setting at high z}.

{\item Once the conditions for a super-critical accretion flow are in place the actual growth rate will depend 
on the gas supply rate to the nuclear envelope/accretion disk as well as on the nature and effectiveness of 
radiative and kinetic feedback from the growing BH. Other effects such as merging with other
black holes in a vigorously star forming CNDs or inside dense nuclear star clusters, might  be important in high-redshift environments.}

\end{itemize}

Future developments will rely heavily on the ability to carry out multi-scale simulations bridging the accretion disk/envelope
domain with the galactic domain at nuclear scales to the very minimum. In parallel, small scale accretion simulations have to investigate
a much wider range of initial conditions, in terms of black hole masses as well as boundary inflow rates and patterns. They also need to quantify
more explicitly the dependence of results, in particular radiative efficiency and mean accretion rate, 
on the adopted radiation transport model. Finally, combining relativistic simulations with the most sophisticated radiation transport methods,
such as VET, is among the technical challenges ahead.

\clearpage

{
\bibliographystyle{ws-rv-har}    
\bibliography{FotFBH_Ch_11}
}


\end{document}